# Inelastic scattering and cooling of photoexcited electrons through coupling with acoustic, optic and surface polar optic phonons in graphene

S. Arshia Khatoon, Meenhaz Ansari, S. S. Z. Ashraf* and M. Obaidurrahman
*Department of Physics, Faculty of Science,*
*Aligarh Muslim University, Aligarh-202002, Uttar Pradesh, India*

Email: *ssz_ashraf@rediffmail.com*

   The acoustic, optic, and surface polar optic phonons are the three important intrinsic and extrinsic phononic modes that increasingly populate graphene on a substrate with rising temperatures; the coupling of which with photoexcited hot carriers in the equipartition regime provides significant pathways for electron-phonon relaxation. In this paper, we theoretically investigate the relative significance of the three phononic modes in electron scattering and cooling phenomena in single layer graphene, including their comparison with supercollision driven power loss, and obtain analytical formulae on the energy dependence of electron-phonon scattering rate and cooling power in the Boltzmann transport formalism. The obtained analytical solutions not only closely reproduce the results for scattering rate and cooling power, as that obtained from the earlier reported numerically tractable integral forms, but also enable us to derive closed-form formulae of the cooling time and thermal conductance. The important role of Pauli blocking that prevents transition to filled energy states has also been elucidated in the estimation of the scattering rate and cooling power density for all the three modes. The obtained formulae provide a better insight into the dynamics of hot electron phenomena giving an explicit view on the interplay of the different variables that affect the transport quantities under investigation. The formulae can also be potentially useful for performance optimization of transport quantities in numerical optimization methods since the first and second-order derivatives are easily deducible from these formulae.

Keywords; Inelastic scattering rate, cooling power, acoustic, optical, surface polar optic phonons, Pauli blocking, cooling time, heat conductance

## I. INTRODUCTION

   The honeycomb structure leading to the generation of linear energy dispersion in graphene is the genesis for the manifestation of many exceptional physical properties including the remarkably distinct optical properties [1]. Because of the unique physical properties, graphene despite being only an atom thick (1.7A0) absorbs a significant fraction (2.3%) of incident light at normal incidence, in contrast to which about 50 atomic layers of silicon (250 A0) are required for similar absorbance in the visible region. The amazing optical properties in tandem with the unique electrical/electronic properties have led graphene to emerge as a prime material on the material landscape for photonic and optoelectronic applications; in as much that the optoelectronic and photonic properties have been touted as the grey area where the true potential for technological applications of graphene really lie [2]. In particular, the gapless energy bandgap in graphene enables efficient interaction with optical waves over a broad absorption spectrum range from terahertz to visible frequencies, and further its compatibility for integration with silicon-based devices makes it an attractive material for optoelectronics and photonics applications [3-4]. The optoelectronic applications are at the heart of several major technologies, including solar cells, light-emitting devices, touch screens, photodetectors, saturable absorbers, ultrafast lasers, optical limiters, optical frequency converters, optical switch, optical communication systems, electro-optic modulators, etc [2-5].

When photons of energy $E_{pt} > hv/2$ are incident on graphene they cause interband optical transitions resulting in the creation of photo excited equal energy sharing electron-hole pairs or photo carriers. In graphene a number of studies have delineated three relaxation time scales one after the other which the photoexcited carriers undergo [6-9]. First an initial fast femtosecond ($t \lesssim 20 \times 10^{-15} sec$) thermalization through carrier-carrier Coulombic interaction occurs, followed by a second slow femto-second($t \lesssim 200 \times 10^{-15} sec$) cooling process of thermalized carriers through emission of optical phonons of energies ($\sim 200 meV$). The third and final process happens through acoustic phonons as the majority of electron distribution have energies less than the optical phonons. In disordered graphene another process named supercollision occurs that ensures the emission of phonons with higher energy and momentum than normal collision, as the momentum conservation constraints are relaxed in supercollision. This supercollision-cooling process is specified by more rapid cooling times and a cubic temperature power dependency in steady state [10-11].

The cooling dynamics of these photocarriers have been the subject of a lot of experimental and theoretical investigations [2-42]. The main relaxation pathways for carriers that have been in contention to dissipate their excess heat have been elucidated as acoustic phonons, optical phonons, super-collision, and surface polar optical phonons in case of graphene on a substrate [10-17]. Each of these mechanisms of energy relaxation has different temperature dependencies and is found to dominate over the others in various electronic temperature-dependent transport regimes as illustrated in Fig.1(a). The transport regime begins from very low-temperature direct heat conduction regime called as Wiedemann-Franz, to inplane acoustic phonon dependent low-temperature Bloch-Gruneisen (BG), to high-temperature supercollision (in case of graphene with the disorder), to very high-temperature non-degenerate electrons equipartition (EP) regime where scattering by intrinsic optical phonons and extrinsic surface polar phonons rule the roost [12,18].

The exact nature and strength of electron-phonon coupling in graphene is still being debated. Specifically, the electron acoustic phonon interaction strength, characterized by the deformation potential (DP) has been controversial [17, 20-21]. A comprehensive understanding of the scattering processes in graphene is vital for determining various transport parameters and quantities like electrical and thermal resistivity/mobility, effective DP, BG temperature limit, ultrafast relaxation processes after photo excitation, cooling rate, optical, remote interfacial and intra-ripple flexural phonon scatterings, and for innumerable direct and indirect applications such as in designing graphene-based bolometric, acousto-electric devices, optoelectronic devices, high-frequency spectrometers, high-quality graphene transistors and strain engineering. [2-26].

Apart from the energy loss rate (P), the electron-phonon (e-p) scattering rate ($\tau^{-1}$) is also an important transport parameter that provides a means to estimate the value of DP besides mobility and other transport quantities. An important consideration in the estimation of e-p scattering rate is the phenomenon of Pauli Blocking (PB) because of which an incoming electron is blocked from making a transition to the final filled transition states [2]. In doped graphene when $E_F > h\nu/2$, only intra-band transitions are allowed since interband transitions are prohibited by PB. The phenomenon of saturable absorption which is so relevant for the construction of saturable absorbers and ultra-fast lasers is very closely connected to PB [2, 24]. When the optical pulse excitation intensity is increased the photo generated carriers concentration increase and begin to exceedingly dominate the intrinsic electron and hole carrier population in graphene. Since the relaxation times are shorter than the pulse duration therefore the states near the edge of the conduction and valence bands are filled prohibiting further absorption. Thus band filling occurs because no two electrons can fill the same state and saturable absorption is achieved due to this PB process.

The PB factor for estimating the scattering rate and other transport properties at low temperatures becomes unity because of the quasi-elastic nature of scattering in the BG regime. But at higher temperatures, that is near and in the EP regime, the phonons are of increasingly higher energy and the quasi-elastic approximation is not justified as the scattering becomes increasingly inelastic and the PB factor begins to influence the scattering rate. Particularly in the case of optical phonon scattering the PB is more relevant as the scattering is wholly inelastic and can rarely be considered elastic. Fig.1(b) is a demonstration of the variation of PB factor with electron energy at room temperature for both longitudinal acoustic and optical phonons. Studies reported earlier for the scattering rate of hot electrons in graphene due to optical and surface polar phonons have neglected the PB factor which is not justified but in the case of acoustic phonons only, where it matters the least[14, 25-29].

In this paper, we revisit and explore these three aspects i.e. electron-phonon relaxation rate, cooling power, and cooling time in the EP regime for single-layer graphene (SLG) and obtain closed-form expressions for the earlier reported integral equations in the Boltzmann transport formalism (BTF). Obviously a closed-form solution potentially gives a better understanding of how different variables affect the transport quantity in question. They are also more useful for performance optimization since one can then easily compute the first and second-order derivatives, which are often used in numerical optimization methods. From the first derivative of the cooling power, we obtain analytical formulae for cooling time and thermal conductance. In this study, the acoustic, optical phonon, and surface polar optical phonon (SOP) scattering have been treated elastically as well as inelastically, and also with and without PB factor. In section II we give the brief formalism and the obtained analytical results followed by a discussion of numerical & analytical results in section III. Finally, we conclude the study in section IV.

**II. FORMALISM AND ANALYTICAL RESULTS**

The e-p interaction affects the characteristic lifetime of electronic states, which is the time for an excited electronic state to decay to the ground state. The energy-dependent scattering rate and cooling power in SLG due to acoustic and optical scattering mechanisms are evaluated using Fermi's golden rule which measures the transition probability $T_{kk'}$ from the energy state with electron momentum $k$ to another state with electron momentum $k'$ as [30],

$$T_{kk'} = \frac{2\pi}{\hbar} \sum_q |\langle k'|H_{e-p}|k\rangle|^2 \, \delta(E_k \pm \hbar\omega_q - E_{k'}) \quad (1)$$

where $\langle k'|H_{e-p}|k\rangle$ is the perturbation matrix element containing the Frohlich Hamiltonian for e-p interaction, $H_{e-p} = \sum_{kq} M_{kq} a^\dagger_{k+q} a_k (b_q + b^\dagger_{-q})$ in which $M_{kq} = M_{kk'}(q) C_{kk'}$ where $M_{kk'}(q)$ is the e-p coupling strength, $C_{kk'} = |C^\dagger_{k+q} C_k|^2 = [1 + \cos\theta_{kk'}]/2$ is the chirality factor that arises from the overlap of wave functions, $a^\dagger_{k+q}(a_k)$ are the electron creation (annihilation) operators, $b_q(b^\dagger_{-q})$ are the phonon annihilation (creation) operators, $\theta$ is the angle between scattering in and out wave vectors $k$ and $k'$. The energy dispersion relation for Dirac electron in graphene is $E_k = \hbar v_F k$, and the Dirac spinor solutions have the form,

$\psi_k(r) = \frac{1}{\sqrt{2A}} \begin{pmatrix} e^{-i\varphi_k/2} \\ e^{i\varphi_k/2} \end{pmatrix} e^{ik.r}$ [31]. Considering that the scattering mechanism involves both the absorption and the emission processes, the square of the matrix element of the e-p interaction evaluates out as $|\langle k'|H_{e-p}|k\rangle|^2 \delta(E_k \pm \hbar\omega_q - E_{k'}) = |M_{kq}|^2 \{N_{\omega_q} \delta(E_k + \hbar\omega_q - E_{k'}) + (N_{\omega_q} + 1)\delta(E_k - \hbar\omega_q - E_{k'})\}$, where $N_{\omega_q} = [\{\exp(\beta\hbar\omega_q) - 1\}]^{-1}$ is the Bose-Einstein distribution function of phonons at lattice temperature $T_l$, where $\beta = 1/k_B T$. The delta functions $\delta(E_k \pm \hbar\omega_q - E_{k'})$ ensure the energy conservation for both inelastic ($\hbar\omega_q \neq 0$) and elastic ($\hbar\omega_q = 0$) scattering processes by the absorption or emission of phonons of energy $\hbar\omega_q$ or wave vector $q = k - k'$. The wave vector $q$ in case of inelastic scattering is equal to $q = \sqrt{k^2 + k'^2 - 2kk'\cos\theta}$ which reduces to $q = 2k \sin(\theta/2)$ for elastic scattering, when, $k = k'$. The square of the matrix element modified by the chirality factor, $|M_{kk'}(q)|^2 C_{kk'}$ due to acoustic, optic and SOP modes in SLG can be written respectively as under [14,26];

$$|M_{kk'}(q)|^2 C_{kk'} = \begin{cases} \dfrac{D_{AP}^2 \hbar \omega_{AP}}{2A\rho v_p^2} \left(\dfrac{1+\cos\theta}{2}\right) & (2) \\ \dfrac{D_{OP}^2 \hbar}{A\rho \omega_{OP}} \left(\dfrac{1+\cos\theta}{2}\right) & (3) \\ \dfrac{D_{SOP}^2 e^2 \exp(-2qd)}{A\, q} \left(\dfrac{1+\cos\theta}{2}\right) & (4) \end{cases}$$

Where $D_{AP}$, $D_{OP}$ and $D_{SOP}^2$ are respectively the e-p potential coupling strength parameter for acoustic, optical and SOP phonons. The SOP e-p coupling parameter is defined as $D_{SOP}^2 = \frac{\hbar\omega_{SOP}}{2}\left(\frac{1}{\varepsilon_{high}+\varepsilon_0} - \frac{1}{\varepsilon_{low}+\varepsilon_0}\right)$, in which $\varepsilon_{high}$ and $\varepsilon_{low}$ are the high and low frequency dielectric permittivity of the substrate, respectively, $\varepsilon_0$ is the free space permittivity, $\omega_{AP}$ & $\omega_{OP}$ are the acoustic and optic frequencies, respectively, $A$ is the area of graphene sheet, $v_p$ is the sound velocity in graphene, $\rho$ is the areal mass density, $e$ is the electronic charge, $d$ is the distance between graphene and substrate. In the subsequent section we briefly discuss the formalism for scattering rate ($\tau^{-1}$), cooling power density ($P$), cooling time ($\tau^c$) & heat conductance ($G$) and obtain closed-form expressions on them for the three phononic modes.

**A. Scattering rate**

The energy dependent scattering rate due to e-p scattering in SLG system can be calculated in the BTF approach as [30],

$$\frac{1}{\tau} = \frac{A}{(2\pi)^2} \int dk'(1 - \cos\theta_{kk'})T_{kk'}\frac{1-f(E_{k'})}{1-f(E_k)} \quad (5)$$

Where the transition probability expands as, $T_{kk'} = \frac{2\pi}{\hbar}|M_{kq}|^2\{N_{\omega_q}\delta(E_k + \hbar\omega_q - E_{k'}) + (N_{\omega_q} + 1)\delta(E_k - \hbar\omega_q - E_{k'})\}$, $f(k) = [\exp\{\beta(E_k - \mu)\} + 1]^{-1}$ is the Fermi-Dirac distribution function, $\mu$ is the chemical potential, $\frac{1-f(E_{k'})}{1-f(E_k)}$ is the PB factor which prevents the transition to filled energy states as a consequence of Pauli exclusion principle. By feeding appropriate matrix elements from Eqs.(2), (3) or (4), one obtains expressions for scattering rate for the three phononic modes, accordingly. For the estimation of scattering rate we consider couplings with only longitudinal acoustic (LA), longitudinal optic (LO) and SOP phonons as other cases can be obtained by just replacing the velocity parameter in the derived prototype solutions.

### 1. Acoustical phonons

Among the e-p interaction processes, we first consider the electron acoustic phonon scattering treated in the frame work of acoustic DP approximation. At sufficiently high temperatures electron-acoustic phonon interaction can be significant especially in intrinsic graphene which can largely inhibit the mobility of electrons [21]. The dispersion relation for the acoustic phonon frequency is considered to be linear, $\omega_q = v_p q$, where $v_p \equiv v_l$ is the longitudinal sound velocity and $q$ is the wave vector associated with the LA phonons. For the evaluation of the energy dependent inelastic electron scattering rate($\tau_{AP}^{-1}$) by LA phonon mode, plugging Eqs.(1) and (2) into Eq. (5) gives the following equation for $\tau_{AP}^{-1}$,

$$\frac{1}{\tau_{AP}} = \frac{D_{AP}^2}{4\pi\rho v_l^2 \hbar^3 v_F^2}\int_0^\infty \frac{1-f(E_{k'})}{1-f(E_k)}E_{k'}dE_{k'}\,\hbar\omega_{AP}\{N_{\omega_{AP}}\delta(E_{k'} - E_k - \hbar\omega_{AP})$$
$$+ (N_{\omega_{AP}} + 1)\delta(E_{k'} - E_k + \hbar\omega_{AP})\}\int_0^{2\pi}\frac{(1-\cos^2\theta)}{2}d\theta \quad (6)$$

Although numerically tractable but in its entirety the above integral is analytically intractable so we approximate, $\hbar\omega_{AP} = \hbar v_l q = \hbar v_l \sqrt{2}k\sqrt{(1-Cos\theta)} \cong \frac{4}{\pi}\left(\frac{v_l}{v_F}\right)E_k$, by taking average of $\langle\sqrt{(1-Cos\theta)}\rangle = \frac{2\sqrt{2}}{\pi}$. Also, in the EP regime we can justifiably approximate $N_{\omega_{AP}} = k_B T/\hbar\omega_{AP}$ as $\hbar\omega_{AP} \ll k_B T$, and by the property of delta function we obtain the following solution within the purview of semi-inelastic electron-acoustic phonon scattering keeping the PB factor,

$$\frac{1}{\tau_{AP}} = \frac{D_{AP}^2 k_B T \left(e^{\frac{\mu}{k_B T}} + e^{\frac{E_k}{k_B T}}\right)}{8\hbar^3 \rho v_l^2 v_F^2 \left(e^{\frac{2\left(1+\frac{2 v_l}{\pi v_F}\right)E_k}{k_B T}} + e^{\frac{E_k+\mu}{k_B T}} + e^{\frac{\left(1+\frac{8 v_l}{\pi v_F}\right)E_k+\mu}{k_B T}} + e^{\frac{\frac{4 v_l}{\pi v_F}E_k+2\mu}{k_B T}}\right)} E_k \times$$

$$\left\{\left(e^{\frac{\frac{8 v_l}{\pi v_F}E_k+\mu}{k_B T}} + e^{\frac{\left(1+\frac{4 v_l}{\pi v_F}\right)E_k}{k_B T}}\right)\left(1+\frac{4}{\pi}\frac{v_l}{v_F}\right) + \left(e^{\frac{\mu}{k_B T}} + e^{\frac{\left(1+\frac{4 v_l}{\pi v_F}\right)E_k}{k_B T}}\right)\left(1-\frac{4}{\pi}\frac{v_l}{v_F}\right)\theta\left(E_k - \frac{4}{\pi}\frac{v_l}{v_F}E_k\right)\right\} \quad (7)$$

where the second term with $\theta$ − Heaviside step function represents phonon emission, which occurs only when electrons with energies greater than $\left(\frac{4}{\pi}\frac{v_l E_k}{v_F}\right)$ are involved. The Eq.(7) expresses the quasi-elastic scattering rate for or an electron in the lowest energy conduction band near the Dirac point for both emission and absorption of acoustic phonons at temperatures higher than the BG temperature. By fitting the calculated velocity/mobility it is possible to extract a value for the acoustic deformation potential [21].

As the LA phonon energy ($\hbar\omega_{AP}$) is quite small as compared to the electronic energy therefore normally the scattering is considered to be quasi-elastic which implies $E_{k'} = E_k$ and $\delta(E_{k'} \pm \hbar\omega_{AP} - E_k) = \delta(E_{k'} - E_k)$, $\frac{1-f(E_{k'})}{1-f(E_k)} = 1$, $\left(\frac{v_l}{v_F}\right) \ll 1$, therefore the above obtained semi-inelastic analytical result reduces to the earlier widely reported quasi-elastic analytical result [21],

$$\frac{1}{\tau_{AP}} = \frac{D_{AP}^2 k_B T}{4\hbar^3 \rho v_l^2 v_F^2} E_k \quad (8)$$

The same result as given in Eq.(8) has been reported in several studies sometimes with different numerical pre-factors for graphene's quasi-elastic scattering rates [21]. From Eq.(8) it can be said that the in plane acoustic phonon scattering is considered to be quasi-elastic when the scattering rate linearly depends both on temperature and electron energy.

## 2. Optical phonons

The scattering by optical phonons is generally an inelastic scattering that is $E_{k'} \neq E_k \Rightarrow f(E_{k'}) \neq f(E_k)$, and hence the PB factor cannot be approximated as unity. As mentioned in the introduction, the earlier reported results have neglected this factor which is justified only for elastic scattering but it cannot be ignored for inelastic scattering as can be noticed from Fig.1(b). We have calculated the optical phonon scattering rate by considering it as an inelastic process and therefore included the role of PB in the estimation of result.

For the evaluation of relaxation rate due to optic phonon mode($\tau_{OP}^{-1}$), substituting Eqs.(1) and (3) into Eq.(5) and proceeding in the same manner, one obtains the following integral equation for $\tau_{OP}^{-1}$ in SLG,

$$\frac{1}{\tau_{OP}} = \frac{D_{OP}^2}{4\pi\rho\omega_{OP}\hbar^2 v_F^2} \int_0^\infty \frac{1-f(E_{k'})}{1-f(E_k)} E_{k'} dE_{k'} \{N_{\omega_{OP}}\delta(E_{k'} - E_k - \hbar\omega_{OP})$$
$$+ (N_{\omega_{OP}} + 1)\delta(E_{k'} - E_k + \hbar\omega_{OP})\} \int_0^{2\pi} \frac{(1-\cos^2\theta)}{2} d\theta \quad (9)$$

Without any approximation, the integration of Eq. (9) yields the following exact solution,

$$\frac{1}{\tau_{OP}} = \frac{D_{OP}^2}{8\rho\omega_{OP}\hbar^2 v_F^2} e^{\frac{E_k}{k_BT}} \left(e^{\frac{-E_k+\mu}{k_BT}} + 1\right) \left( \frac{e^{\frac{\hbar\omega_{OP}}{k_BT}}(E_k + \hbar\omega_{OP})N_{\omega_{OP}}}{\left(e^{\frac{\mu}{k_BT}} + e^{\frac{E_k+\hbar\omega_{OP}}{k_BT}}\right)} \right.$$
$$\left. + \frac{(E_k - \hbar\omega_{OP})(N_{\omega_{OP}} + 1)\theta(E_k - \hbar\omega_{OP})}{\left(e^{\frac{\mu+\hbar\omega_{OP}}{k_BT}} + e^{\frac{E_k}{k_BT}}\right)} \right) \quad (10)$$

The Eq.(10) represents the complete unscreened inelastic scattering rate closed form solution for the electron-optical phonon interaction via optical DP coupling in SLG. It is to be noted that on ignoring the PB factor, $\left(\frac{1-f(E_{k'})}{1-f(E_k)}\right) \approx 1$, the relaxation rate reduces to the earlier reported result [21,33];

$$\frac{1}{\tau_{OP}} = \frac{D_{OP}^2}{8\rho\omega_{OP}\hbar^2 v_F^2} \{(E_k + \hbar\omega_{OP})N_{\omega_{OP}} + (E_k - \hbar\omega_{OP})(N_{\omega_{OP}} + 1)\theta(E_k - \hbar\omega_{OP})\} \quad (11)$$

## 3. Surface polar optical phonons

The transport in graphene placed on a substrate is also affected by the remote phonons that arise from the polar nature of the substrate. The polar substrate creates SOP modes which can strongly scatter the electrons and limit the mobility [12, 26-29]. The SOP scattering rate can be estimated from Eq.(5) using Eqs. (1) & (4) as [28],

$$\frac{1}{\tau_{SOP}} = \frac{{D_{SOP}^i}^2 \pi e^2}{\epsilon_0 \hbar^2 v_F^2} \int_0^\infty \frac{1-f(E_{k'})}{1-f(E_k)} E_{k'} dE_{k'} \left\{ N_{\omega_{SOP}^i} \delta(E_{k'} - E_k - \hbar\omega_{SOP}^i) \right.$$

$$\left. + \left(N_{\omega_{SOP}^i} + 1\right) \delta(E_{k'} - E_k + \hbar\omega_{SOP}^i) \right\} \int_0^{2\pi} \frac{e^{-2q \cdot d}}{q} \left(\frac{1+\cos\theta}{2}\right) d\theta \quad (12)$$

Where the index $i$ stands for different phonon modes of energies $\hbar\omega^i{}_{SOP}$, and here $q = \sqrt{k^2 + {k'}^2 - 2kk'\cos\theta}$. Typically $e^{-2q \cdot d} \cong 1$ for distances upto $\sim 5 A^0$ between graphene and substrate, and without the PB factor the Eq.(12) yields a closed form solution as under;

$$\frac{1}{\tau_{SOP}} = \frac{{D_{SOP}^i}^2 e^2}{24 \hbar^2 \epsilon_0 v_F E_k^2} \left\{ \frac{(1+N_{\omega_{SOP}^i})}{(E_k - \hbar\omega_{SOP}^i)} \left( \sqrt{(-2E_k + \hbar\omega_{SOP}^i)^2} \left( 2E_k^2 - 2E_k \hbar\omega_{SOP}^i + \right. \right. \right.$$

$$\left. \hbar^2 {\omega_{SOP}^i}^2 \right) \text{EllipticE}\left[\frac{4E_k(E_k - \hbar\omega_{SOP}^i)}{(-2E_k + \hbar\omega_{SOP}^i)^2}\right] - \hbar^2 {\omega_{SOP}^i}^2 \text{EllipticK}\left[\frac{4E_k(E_k - \hbar\omega_{SOP}^i)}{(-2E_k + \hbar\omega_{SOP}^i)^2}\right] \right) + \hbar\omega_{SOP}^i \left( \left(2E_k^2 - 2E_k \hbar\omega_{SOP}^i + \right. \right.$$

$$\left. \hbar^2 {\omega_{SOP}^i}^2 \right) \text{EllipticE}\left[\frac{4E_k(-E_k + \hbar\omega_{SOP}^i)}{\hbar^2 {\omega_{SOP}^i}^2}\right] - (-2E_k + \hbar\omega_{SOP}^i)^2 \text{EllipticK}\left[\frac{4E_k(-E_k + \hbar\omega_{SOP}^i)}{\hbar^2 {\omega_{SOP}^i}^2}\right] \right) \theta[E_k - \hbar\omega_{SOP}^i] +$$

$$\frac{1}{E_k + \hbar\omega_{SOP}^i} \left( \hbar\omega_{SOP}^i \left( \left(2E_k^2 + 2E_k \hbar\omega_{SOP}^i + \hbar^2 {\omega_{SOP}^i}^2 \right) \text{EllipticE}\left[-\frac{4E_k(E_k + \hbar\omega_{SOP}^i)}{\hbar^2 {\omega_{SOP}^i}^2}\right] - (2E_k + \right.\right.$$

$$\left. \hbar\omega_{SOP}^i)^2 \text{EllipticK}\left[-\frac{4E_k(E_k + \hbar\omega_{SOP}^i)}{\hbar^2 {\omega_{SOP}^i}^2}\right] \right) + \sqrt{(2E_k + \hbar\omega_{SOP}^i)^2} \left( 2E_k^2 + 2E_k \hbar\omega_{SOP}^i + \right.$$

$$\left. \left. \hbar^2 {\omega_{SOP}^i}^2 \right) \text{EllipticE}\left[\frac{4E_k(E_k + \hbar\omega_{SOP}^i)}{(2E_k + \hbar\omega_{SOP}^i)^2}\right] - \hbar^2 {\omega_{SOP}^i}^2 \text{EllipticK}\left[\frac{4E_k(E_k + \hbar\omega_{SOP}^i)}{(2E_k + \hbar\omega_{SOP}^i)^2}\right] \right) \theta[E_k + \hbar\omega_{SOP}^i] \right\} \quad (13)$$

Where EllipticK and EllipticE are the elliptic functions of first and second kinds respectively. The inclusion of the PB factor modifies the above solution to;

$$\frac{1}{\tau_{SOP}} = \frac{{D_{SOP}^i}^2 e^2}{24 \hbar^2 \epsilon_0 v_F E_k^2} \cdot \frac{e^{\frac{E_k + \hbar\omega_{SOP}^i}{k_B T}} (1 + e^{\frac{-E_k + \mu}{k_B T}})}{\left(-1 + e^{\frac{\hbar\omega_{SOP}^i}{k_B T}}\right)} \left\{ \frac{1}{(e^{\frac{E_k}{k_B T}} + e^{\frac{\mu + \hbar\omega_{SOP}^i}{k_B T}})(E_k - \hbar\omega_{SOP}^i)} \left( \sqrt{(-2E_k + \hbar\omega_{SOP}^i)^2} \left( 2E_k^2 - \right. \right. \right.$$

$$\left. 2E_k \hbar\omega_{SOP}^i + \hbar^2 {\omega_{SOP}^i}^2 \right) \text{EllipticE}\left[\frac{4E_k(E_k - \hbar\omega_{SOP}^i)}{(-2E_k + \hbar\omega_{SOP}^i)^2}\right] - \hbar^2 {\omega_{SOP}^i}^2 \text{EllipticK}\left[\frac{4E_k(E_k - \hbar\omega_{SOP}^i)}{(-2E_k + \hbar\omega_{SOP}^i)^2}\right] \right) +$$

$$\hbar\omega_{SOP}^i \left( \left(2E_k^2 - 2E_k \hbar\omega_{SOP}^i + \hbar^2 {\omega_{SOP}^i}^2 \right) \text{EllipticE}\left[\frac{4E_k(-E_k + \hbar\omega_{SOP}^i)}{\hbar^2 {\omega_{SOP}^i}^2}\right] - (-2E_k + \right.$$

$$\left. \hbar\omega_{SOP}^i)^2 \text{EllipticK}\left[\frac{4E_k(-E_k + \hbar\omega_{SOP}^i)}{\hbar^2 {\omega_{SOP}^i}^2}\right] \right) \theta[E_k - \hbar\omega_{SOP}^i] + \frac{1}{(e^{\frac{\mu}{k_B T e}} + e^{\frac{E_k + \hbar\omega_{SOP}^i}{k_B T e}})(E_k + \hbar\omega_{SOP}^i)} \left( \hbar\omega_{SOP}^i \left( \left(2E_k^2 + \right. \right. \right.$$

$$2E_k\hbar\omega_{SOP}^i + \hbar^2{\omega_{SOP}^i}^2\bigg)\text{EllipticE}\left[-\frac{4E_k\left(E_k+\hbar\omega_{SOP}^i\right)}{\hbar^2{\omega_{SOP}^i}^2}\right] - \left(2E_k + \hbar\omega_{SOP}^i\right)^2\text{EllipticK}\left[-\frac{4E_k\left(E_k+\hbar\omega_{SOP}^i\right)}{\hbar^2{\omega_{SOP}^i}^2}\right]\bigg) +$$

$$\sqrt{\left(2E_k + \hbar\omega_{SOP}^i\right)^2}\bigg(\left(2E_k^2 + 2E_k\hbar\omega_{SOP}^i + \hbar^2{\omega_{SOP}^i}\right)\text{EllipticE}\left[\frac{4E_k\left(E_k+\hbar\omega_{SOP}^i\right)}{\left(2E_k+\hbar\omega_{SOP}^i\right)^2}\right] -$$

$$\hbar^2{\omega_{SOP}^i}^2\text{EllipticK}\left[\frac{4E_k\left(E_k+\hbar\omega_{SOP}^i\right)}{\left(2E_k+\hbar\omega_{SOP}^i\right)^2}\right]\bigg)\bigg)\theta[E_k + \hbar\omega_{SOP}^i]\bigg\} \quad (14)$$

Unlike the case of acoustic and optical phonon scattering where restricted analytical results given by Eq.(8) and (11) exist in the literature, but no such closed form solutions have been reported for the case of SOP scattering, and the governing BTF integral Eq.(12) has only been numerically solved under approximations [14, 21, 27]. All the obtained analytical results on scattering rate in graphene with and without PB and the corresponding reported results for semiconductor from Ref.[34] have been summarized in the Table I.

**B. Cooling power**

The cooling occurs due to transfer of energy from the electron bath to the lattice and the cooling or energy loss power ($P$) measures the rate at which the hot-electron distribution loses its energy to the lattice. The cooling power per unit area for graphene is given by,

$$P = -\frac{g_s g_v}{A}\sum_{k,k'}(E_{k'} - E_k)\left(\frac{\partial N_{\omega q}}{\partial t}\right)_{Coll.} \quad (15)$$

where $g_s$ & $g_v$ are spin and valley degeneracies, $E_{k'} = E_k + \hbar\omega_q$, and $\left(\frac{\partial N_{\omega q}}{\partial t}\right)$ is the collision integral that describes the variation rate of the phonon distribution function $N_{\omega q}$ by e–p scattering [14]. On feeding the value of collision integral the function $P$ reduces to [15],

$$P = -\frac{g_s g_v}{A}\sum_{k,k'}(E_{k'} - E_k)T_{kk'}f(E_k)(1 - f(E_{k'})) \quad (16)$$

From this Eq.(15) we calculate in the subsequent sub sections the cooling power due to electron acoustic, optical and SOP phonons coupling respectively and obtain analytical results on them. We follow the formalism presented in Ref.[15] for obtaining the cooling power.

1. **Acoustical phonons**

In the case of cooling power due to acoustic phonons the interband transitions are forbidden because of energy-momentum conservations. Using Eqs.(1) and (2), the Eq.(16) for the intra-conduction band cooling power ($P_{AP}$) transforms to [15],

$$P_{AP} = -\frac{g_s g_v}{\hbar A}\sum_{k,k'}|M_{kk'}(q)|^2 C_{kk'}\delta(E_{k'} - E_k - \hbar\omega_{AP})(E_{k'} - E_k)S_{kk'} \quad (17)$$

Where, $|M_{kk'}(q)|^2$ is the total matrix elements due to combined two acoustic phonon modes of LA and TA in which the chirality factor $\left(\frac{1+\cos\theta}{2}\right)$ in the conduction band has been negated, because of the fact that the LA and TA modes have different angular dependencies, and $S_{kk'} = (N_{\omega_{AP}} + 1)f(E_{k'})(1 - f(E_k)) - N_{\omega_{AP}}f(E_k)(1 - f(E_{k'}))$ can be approximated as [15],

$$S_{kk'} = N_{\omega_{AP};T_l}\left(e^{\frac{\hbar\omega_{AP}}{k_B T_l}} - e^{\frac{\hbar\omega_{AP}}{k_B T_e}}\right)f(E_{k'})(1 - f(E_k)) \quad (18)$$

With the help of Eq.(18), the Eq.(17) takes the form [15],

$$P_{AP} = -\frac{g_s g_v}{(2\pi)^2}\frac{D_{AP}^2}{2\rho\hbar^3 v_F^4}\int_0^\infty E_k(E_k + \hbar\omega_{AP})dE_k \int_0^{2\pi} q^2 d\theta\, N_{\omega_{AP};T_l}\left(e^{\frac{\hbar\omega_{AP}}{k_B T_l}} - e^{\frac{\hbar\omega_{AP}}{k_B T_e}}\right)f(E_{k'})(1 - f(E_k)) \quad (19)$$

As, $\hbar\omega_{AP} \ll k_B T_l, k_B T_e$, therefore $N_{\omega_{AP};T_l}\left(e^{\frac{\hbar\omega_{AP}}{k_B T_l}} - e^{\frac{\hbar\omega_{AP}}{k_B T_e}}\right)f(E_{k'})(1 - f(E_k)) \approx f(E_k)(1 - f(E_k))\left(\frac{T_e - T_l}{T_e}\right)$. The integral equation is still not amenable to analytical evaluation, hence to extract an analytical solution we again do the same approximation as done in the case of scattering rate by acoustical phonons and obtain the following equation,

$$P_{AP} \cong \int_0^\infty \frac{16\, D_{AP}^2}{\pi^3 \rho\hbar^5 v_F^6}\frac{e^{\frac{-\mu+E_k\left(1+\frac{4v_p}{\pi v_F}\right)}{k_B T_e}} E_k^4\left(1+\frac{4v_p}{\pi v_F}\right)dE_k}{\left(1+e^{\frac{-\mu+E_k}{k_B T_e}}\right)\left(1+e^{\frac{-\mu+E_k\left(1+\frac{4v_p}{\pi v_F}\right)}{k_B T_e}}\right)}\left(\frac{T_e - T_l}{T_e}\right) \quad (20)$$

Further we approximate, $f(E_k)(1 - f(E_k)) = f(E_k) - (f(E_k))^2\, (f(E_k))^2$ as $f(E_k) - e^{\frac{(\mu - E_k)}{k_B T}}$ in, as $e^{\frac{(E_k - \mu)}{k_B T}} > 1$ for $T \leq 1000K$ and $\mu \leq 0.2\, eV$, so Eq.(20) becomes,

$$P_{AP} \cong \int_0^\infty \frac{16\, D_{AP}^2}{\pi^3 \rho\hbar^5 v_F^6}\left(\frac{1}{1+e^{\frac{-\mu+E_k}{k_B T_e}}} - e^{\frac{2(\mu - E_k)}{k_B T_e}}\right)E_k^4\left(1 + \frac{4}{\pi}\frac{v_p}{v_F}\right)dE_k\left(\frac{T_e - T_L}{T_e}\right) \quad (21)$$

On integrating Eq.(21) yields the following solution,

$$P_{AP} \cong -\frac{48\, D_{AP}^2 K_B^5 T_e^5}{\pi^3\, \rho v_F^6 \hbar^5}\left(\frac{\Delta T}{T_e}\right)\left(1 + \frac{4}{\pi}\frac{v_p}{v_F}\right)\left(8\text{PolyLog}[5, -e^{\frac{\mu}{k_B T_e}}] + \frac{1}{4}e^{\frac{\mu}{k_B T_e}}\right) \quad (22)$$

The Eq.(22) is the nearly approximate analytical result of Eq.(19), and Eq.(19) represents the complete numerical integral equation for the average intraband electron cooling rate through acoustic phonons via DP coupling in SLG in EP regime. For $\mu = 0$ the Eq.(22) reduces to $P_{AP} = -\frac{48 D_{AP}^2 K_B^5 T_e^5}{\pi^3 \rho v_F^6 \hbar^5}\left(\frac{\Delta T}{T_e}\right)\left(1 + \frac{4}{\pi}\frac{v_p}{v_F}\right)$ where $\Delta T = T_e - T_l$. The Eq.(21) can be compared with the simplified unsolved integral form given in Ref. [15] which is quoted below for ease of reference,

$$P_{AP;TL} \approx -\frac{g_s g_v}{2\pi}\frac{D_{AP}^2}{\rho v_F}\frac{T_e - T_l}{T_e}\int_0^\infty k^4 dk\, f(E_k)(1 - f(E_k)) \quad (23)$$

The above Eq.(23) is reported in Ref.[15] with a typographic error, that is $v_s$ occurs instead of $v_F$.

### 1. Optical phonons

The cooling power ($P_{OP}$) using Eqs.(1), (3) and Eq.(16), due to combined LO and TO phonon modes intra-conduction band scattering in the BTF framework takes the form [15],

$$P_{OP} = -\frac{g_s g_v}{(2\pi)^2 \hbar^5 v_F^4}\int_0^\infty E_k dE_k \int_0^\infty E_{k'} dE_{k'} \int_0^{2\pi} d\theta\, \frac{D_{OP}^2 \hbar}{2\rho\omega_0}\delta(E_{k'} - E_k - \hbar\omega_{OP})(E_{k'} - E_k) S_{kk'} \quad (24)$$

in which the chirality factor has been negated, for similar reasons like in the case of LA & TA phonon scattering. On evaluating the delta integral the above Eq.(24) reduces to the following integral equation,

$$P_{OP} = -\frac{g_s g_v D_{OP}^2}{8\pi^2 \rho \hbar^3 v_F^4}\int_0^\infty E_k(E_k + \hbar\omega_{OP})dE_k \int_0^{2\pi} d\theta\, S_{kk'} \quad (25)$$

Plugging the value of $S_{kk'} = N_{\omega_{OP};T_l}\left(e^{\frac{\hbar\omega_{OP}}{k_B T_l}} - e^{\frac{\hbar\omega_{OP}}{k_B T_e}}\right)f(E_{k'})(1 - f(E_k))$ into Eq.(25), one obtains the following integral equation as reported in Ref. [15],

$$P_{OP} = -\frac{g_s g_v D_{OP}^2}{4\pi\rho\hbar^3 v_F^4}N_{\omega_{OP}}\left(e^{\frac{\hbar\omega_{OP}}{k_B T_l}} - e^{\frac{\hbar\omega_{OP}}{k_B T_e}}\right)\int_0^\infty \frac{e^{\frac{-\mu+E_k}{k_B T_e}} E_k(E_k + \hbar\omega_{OP})dE_k}{\left(1 + e^{\frac{-\mu+E_k}{k_B T_e}}\right)\left(1 + e^{\frac{-\mu+E_k+\hbar\omega_{OP}}{k_B T_e}}\right)} \quad (26)$$

The Eq.(26) is also unwieldy analytically, therefore making another approximation, $\left(1 + e^{\frac{-\mu+E_k+\hbar\omega_{OP}}{k_B T_e}}\right) = e^{\frac{-\mu+E_k+\hbar\omega_{OP}}{k_B T_e}}$ as, $e^{\frac{-\mu+E_k+\hbar\omega_{OP}}{k_B T_e}} \gg 1$ the above equation becomes,

$$P_{OP} \approx -\frac{D_{OP}^2}{\pi\rho\hbar^3 v_F^4} N_{\omega_{OP}} \left(e^{\frac{\hbar\omega_{OP}}{k_B T_l}} - e^{\frac{\hbar\omega_{OP}}{k_B T_e}}\right) \int_0^\infty \frac{e^{\frac{-\hbar\omega_{OP}}{k_B T_e}} E_k(E_k+\hbar\omega_{OP}) dE_k}{\left(1+e^{\frac{-\mu+E_k}{k_B T_e}}\right)} \quad (27)$$

This Eq.(27) yields a closed form solution,

$$P_{OP} = -\frac{D_{OP}^2 k_B^2 T_e^2}{6\pi\rho\hbar^3 v_F^4} N_{\omega_{OP}} \left(e^{\frac{\hbar\omega_{OP}}{k_B T_l}} - e^{\frac{\hbar\omega_{OP}}{k_B T_e}}\right) e^{\frac{-\hbar\omega_{OP}}{k_B T_e}} \left\{2k_B T_e \text{PolyLog}[3, -e^{\frac{\mu}{k_B T_e}}] + \hbar\omega_{OP} \text{PolyLog}[2, -e^{\frac{\mu}{k_B T_e}}]\right\} \quad (28)$$

The Eq. (28) is the nearly close analytical result of Eq.(26), and the Eq.(26) represents the complete numerical integral equation for the average intraband cooling rate through optical DP coupling in SLG in EP regime. The Eq. (28) can be compared with the analytical solution given as in Ref. [8-9];

$$P_{OP} = -\frac{D_{OP}^2 (\hbar\omega_{OP})^4}{12\hbar^3 v_F^2 E_F} (N_{\omega_{OP};T_l} - N_{\omega_{OP};T_e}) f(-\mu) \quad (29)$$

## 2. Surface polar optical phonons

The cooling rate due to intraband inelastic electron- multiple SOP mode scattering in the BTF framework using Eqs.(1) and (4) and Eq.(16) takes the form [15],

$$P_{SOP} = -\frac{g_s g_v}{(2\pi)^2 \hbar^5 v_F^4} \int_0^\infty E_k dE_k \int_0^\infty E_{k'} dE_{k'} \int_0^{2\pi} d\theta \frac{{D_{SOP}^i}^2 \pi e^2 e^{-2qd}}{\epsilon_0 q} \left(\frac{1+\cos\theta}{2}\right) \delta(E_{k'} - E_k - \hbar\omega^i{}_{SOP})(E_{k'} - E_k) S_{kk'}^i \quad (30)$$

Where the index $i$ stands for different phonon modes of energies $\hbar\omega^i{}_{SOP}$. On evaluating the delta integral the above Eq.(30) reduces to the following integral equation,

$$P_{SOP} = -\frac{g_s g_v}{(2\pi)^2 \hbar^5 v_F^4} \int_0^\infty E_k (E_k + \hbar\omega_{SOP}^i) dE_k \int_0^{2\pi} d\theta \frac{{D_{SOP}^i}^2 \pi e^2 e^{-2qd}}{\epsilon_0 q} \left(\frac{1+\cos\theta}{2}\right) S_{kk'}^i \quad (31)$$

Where, $S_{kk'}^i = N_{\omega_{SOP}^i} \left(e^{\frac{\hbar\omega_{SOP}^i}{k_B T_l}} - e^{\frac{\hbar\omega_{SOP}^i}{k_B T_e}}\right) f(E_{k'})(1 - f(E_k))$. Using this the Eq.(31) reduces to,

$$P_{SOP} = \frac{{D_{SOP}^i}^2 e^2 \omega_{SOP}^i}{\pi\epsilon_0 \hbar^4 v_F^4} N_{\omega_{SOP}^i} \left( e^{\frac{\hbar\omega_{SOP}^i}{k_B T_l}} - e^{\frac{\hbar\omega_{SOP}^i}{k_B T_e}} \right) \int_0^\infty E_k (E_k +$$

$$\hbar\omega_{SOP}^i) dE_k \int_0^{2\pi} d\theta \frac{e^{-2qd}}{q} \left(\frac{1+\cos}{2}\right) f(E_{k'})(1-f(E_k)) \quad (32)$$

On performing $\theta$-integration one gets,

$$P_{SOP} = \frac{{D_{SOP}^i}^2 e^2 \omega_{SOP}^i}{\epsilon_0 \hbar^4 v_F^4} N_{\omega_{SOP}^i} \left( e^{\frac{\hbar\omega_{SOP}^i}{k_B T_l}} - e^{\frac{\hbar\omega_{SOP}^i}{k_B T_e}} \right) \int_0^\infty \frac{e^{-2q}}{q} \frac{e^{\frac{-\mu+E_k}{k_B T_e}} E_k(E_k+\hbar\omega_{SOP}^i) dE_k}{\left(1+e^{\frac{-\mu+E_k}{k_B T_e}}\right)\left(1+e^{\frac{-\mu+E_k+\hbar\omega_{SOP}^i}{k_B T_e}}\right)} \quad (33)$$

This integral is also is not analytically solvable hence to enable a solution of this integral equation we justifiably approximate $e^{-2qd} = 1$,

$$q = \sqrt{k^2 + \left(k + \frac{\omega_{SOP}^i}{v_F}\right)^2 - 2k\left(k + \frac{\omega_{SOP}^i}{v_F}\right)\cos\theta} \cong \left(k + \frac{\omega_{SOP}^i}{v_F}\right) \text{ and } f(E_{k'}) \cong \frac{1}{e^{\frac{-\mu+E_k+\hbar\omega_{SOP}^i}{k_B T_e}}} \text{ as}$$

$e^{\frac{-\mu+E_k+\hbar\omega_{SOP}^i}{k_B T_e}} > 1$ for $T \leq 1000K$ and $\mu \leq 0.2\ eV$. Making these approximations yields the following integral equation,

$$P_{SOP} = \frac{{D_{SOP}^i}^2 e^2 \omega_{SOP}^i}{\epsilon_0 \hbar^3 v_F^3} N_{\omega_{SOP}^i} \left( e^{\frac{\hbar\omega_{SOP}^i}{k_B T_l}} - e^{\frac{\hbar\omega_{SOP}^i}{k_B T_e}} \right) \int_0^\infty \frac{e^{\frac{-\hbar\omega_{SOP}^i}{k_B T_e}} E_k dE_k}{\left(1+e^{\frac{-\mu+E_k}{k_B T_e}}\right)} \quad (34)$$

The Eq.(34) on integration yields the analytical result as under,

$$P_{SOP} = \frac{{D_{SOP}^i}^2 e^2 \omega_{SOP}^i k_B^2 T_e^2}{\epsilon_0 \hbar^3 v_F^3} \frac{e^{-\frac{\hbar\omega_{SOP}^i}{k_B T_e}} \left( e^{\frac{\hbar\omega_{SOP}^i}{k_B T_e}} - e^{\frac{\hbar\omega_{SOP}^i}{k_B T_l}} \right) \text{PolyLog}[2, -e^{\frac{\mu}{k_B T_e}}]}{\left( e^{\frac{\hbar\omega_{SOP}^i}{k_B T_l}} - 1 \right)} \quad (35)$$

The Eq.(35) is the nearly approximate analytical result of Eq.(33) which represents the complete numerical integral equation for the average intraband cooling rate through multiple SOP DP couplings in SLG in EP regime. In Table II the analytical formulae obtained for cooling power along with cooling time and heat conductance have been listed.

| Phonon Mode | Inelastic Scattering Rate in Graphene with PB $(\tau_{AP/OP/SOP})^{-1}$ | Inelastic Scattering Rate in Graphene without PB $(\tau_{AP/OP/SOP})^{-1}$ | *Elastic/Inelastic Scattering Rate in Semiconductors without PB $(\tau_{AP/OP/SOP})^{-1}$ |
|---|---|---|---|
| Acoustic Phonons | $\dfrac{D_{AP}{}^2 k_B T \left(e^{\frac{\mu}{k_B T}}+e^{\frac{E_k}{k_B T}}\right)E_k}{8\hbar^3 \rho v_p{}^2 v_F{}^2 \left(e^{\frac{2\left(1+\frac{2v_p}{\pi v_F}\right)E_k}{k_B T}}+e^{\frac{E_k+\mu}{k_B T}}+e^{\frac{\left(1+\frac{8v_p}{\pi v_F}\right)E_k+\mu}{k_B T}}+e^{\frac{\frac{4v_p}{\pi v_F}E_k+2\mu}{k_B T}}\right)}$ $\times\left\{\left(e^{\frac{\frac{8v_p}{\pi v_F}E_k+\mu}{k_B T}}+e^{\frac{\left(1+\frac{4v_p}{\pi v_F}\right)E_k}{k_B T}}\right)\left(1+\frac{4}{\pi}\frac{v_p}{v_F}\right)\right.$ $\left.+\left(e^{\frac{\mu}{k_B T}}+e^{\frac{\left(1+\frac{4v_p}{\pi v_F}\right)E_k}{k_B T}}\right)\left(1-\frac{4}{\pi}\frac{v_p}{v_F}\right)\theta\left(E_k-\frac{4}{\pi}\frac{v_p}{v_F}E_k\right)\right\}$ | $\dfrac{D_{AP}{}^2 k_B T}{8\hbar^3 \rho v_p{}^2 v_F{}^2}E_k$ $\times\left\{\left(1+\frac{4}{\pi}\frac{v_p}{v_F}\right)\right.$ $\left.+\left(1-\frac{4}{\pi}\frac{v_p}{v_F}\right)\theta\left(E_k\left(1-\frac{4}{\pi}\frac{v_p}{v_F}\right)\right)\right\}$ | $\dfrac{D_{AP}{}^2 (2m_e)^{3/2} k_B T_e}{2\pi \hbar^4 \rho v_p{}^2}\sqrt{E_k}$ |
| Optical Phonons | $\dfrac{D_{OP}{}^2}{8\rho \omega_{OP}\hbar^2 v_F{}^2}e^{\frac{E_k}{k_B T}}\left(e^{\frac{-E_k+\mu}{k_B T}}+1\right)$ $\times\left(\dfrac{e^{\frac{\hbar\omega_{OP}}{k_B T}}(E_k+\hbar\omega_{OP})N_{\omega_{OP}}}{\left(e^{\frac{\mu}{k_B T}}+e^{\frac{E_k+\hbar\omega_{OP}}{k_B T}}\right)}\right.$ $\left.+\dfrac{(E_k-\hbar\omega_{OP})(N_{\omega_{OP}}+1)\theta(E_k-\hbar\omega_{OP})}{\left(e^{\frac{\mu+\hbar\omega_{OP}}{k_B T}}+e^{\frac{E_k}{k_B T}}\right)}\right)$ | $\left(\dfrac{D_{OP}{}^2}{8\rho\omega_{OP}\hbar^2 v_F{}^2}\right)\times$ $\{(E_k+\hbar\omega_{OP})N_{\omega_{OP}}\theta(E_k+\hbar\omega_{OP})$ $+(E_k-\hbar\omega_{OP})$ $(N_{\omega_{OP}}+1)\theta(E_k-\hbar\omega_{OP})\}$ | $\dfrac{D_{OP}{}^2(2m_e)^{3/2}}{4\pi\hbar^3 \rho\omega_{OP}}\times$ $\{\sqrt{(E_k+\hbar\omega_{OP})}N_{\omega_{OP}}\theta(E_k+\hbar\omega_{OP})$ $+\sqrt{(E_k-\hbar\omega_{OP})}$ $(N_{\omega_{OP}}+1)\theta(E_k-\hbar\omega_{OP})\}$ |
| Surface Optical Phonons | $\dfrac{D_{SOP}^i{}^2 e^2}{24\,\epsilon_0 v_F \hbar^2 E_k{}^2}N_{\omega_{SOP}^i}e^{\frac{E_k+\hbar\omega_{SOP}^i}{k_B T}}\left(1+e^{\frac{-E_k+\mu}{k_B T}}\right)$ $\times\left\{\dfrac{1}{(e^{\frac{E_k}{k_B T}}+e^{\frac{\mu+\hbar\omega_{SOP}^i}{k_B T}})t_1}\left(t_2\left(t_3\ \text{EllipticE}\left[\frac{t_4}{t_2{}^2}\right]\right.\right.\right.$ $\left.-\hbar^2\omega_{SOP}^i{}^2\,\text{EllipticK}\left[\frac{t_4}{t_2{}^2}\right]\right)$ $+\hbar\omega_{SOP}^i\left(t_3\text{EllipticE}\left[\frac{-t_4}{\hbar^2\omega_{SOP}^i{}^2}\right]\right.$ $\left.\left.-t_2{}^2\text{EllipticK}\left[\frac{-t_4}{\hbar^2\omega_{SOP}^i{}^2}\right]\right)\right)\theta[t_1]$ $+\dfrac{1}{(e^{\frac{\mu}{k_B T}}+e^{\frac{E_k+\hbar\omega_{SOP}^i}{k_B T}})t_5}\left(\hbar\omega_{SOP}^i\left(t_6\text{EllipticE}\left[-\frac{t_7}{\hbar^2\omega_{SOP}^i{}^2}\right]\right.\right.$ $\left.-t_2{}^2\text{EllipticK}\left[-\frac{t_7}{\hbar^2\omega_{SOP}^i{}^2}\right]\right)$ $+t_8\left(t_6\text{EllipticE}\left[\frac{t_7}{t_8{}^2}\right]\right.$ $\left.\left.\left.-\hbar^2\omega_{SOP}^i{}^2\text{EllipticK}\left[\frac{t_7}{t_8{}^2}\right]\right)\right)\theta[t_5]\right\}$ | $\dfrac{D_{SOP}^i{}^2 e^2}{24\hbar^2\epsilon_0 v_F E_k{}^2}\left\{\dfrac{(1+N_{\omega_{SOP}^i})}{t_1}\right.$ $\times\left(t_2\left(t_3\text{EllipticE}\left[\frac{t_4}{t_2{}^2}\right]\right.\right.$ $\left.-\hbar^2\omega_{SOP}^i{}^2\text{EllipticK}\left[\frac{t_4}{t_2{}^2}\right]\right)$ $+\hbar\omega_{SOP}^i\left(t_3\text{EllipticE}\left[\frac{-t_4}{\hbar^2\omega_{SOP}^i{}^2}\right]\right.$ $\left.\left.-t_2{}^2\text{EllipticK}\left[\frac{-t_4}{\hbar^2\omega_{SOP}^i{}^2}\right]\right)\right)\theta[t_1]$ $+\dfrac{1}{t_5}\left(\hbar\omega_{SOP}^i\left(t_6\text{EllipticE}\left[-\frac{t_7}{\hbar^2\omega_{SOP}^i}\right]\right.\right.$ $\left.-t_2{}^2\text{EllipticK}\left[-\frac{t_7}{\hbar^2\omega_{SOP}^i{}^2}\right]\right)$ $+t_8\left(t_6\text{EllipticE}\left[\frac{t_7}{t_8{}^2}\right]\right.$ $\left.\left.\left.-\hbar^2\omega_{SOP}^i{}^2\text{EllipticK}\left[\frac{t_7}{t_8{}^2}\right]\right)\right)\theta[t_5]\right\}$ | $\dfrac{\omega_{SOP}^i\left(\frac{1}{\epsilon_\infty}-\frac{1}{\epsilon_0}\right)e^2}{2\pi\hbar\epsilon_\infty\sqrt{\frac{2E_k}{m_e}}}\times$ $\left\{\text{Sinh}^{-1}\sqrt{\left(\frac{E_k}{\hbar\omega_{SOP}^i}\right)}N_{\omega_{SOP}^i}\theta\left(\frac{E_k}{\hbar\omega_{SOP}^i}\right)\right.$ $+\text{Sinh}^{-1}\sqrt{\left(\frac{E_k}{\hbar\omega_{SOP}^i}-1\right)}$ $\left.(N_{\omega_{SOP}^i}+1)\theta\left(\frac{E_k}{\hbar\omega_{SOP}^i}-1\right)\right\}$ |

TABLE I. Summary of analytical results for scattering rate, with and without PB, through acoustic, optic and SOP scattering in SLG and Semiconductors. The optic without PB rate result is from Ref.[21] and the all the semiconductor scattering rate formulae are from Ref.[34]. The new symbols introduced in the $(\tau_{SOP})^{-1}$ formulae for compactness are defined as; $t_1=(E_k-\hbar\omega_{SOP}^i), t_2=(-2E_k+\hbar\omega_{SOP}^i), t_3=(2E_k{}^2-2E_k\hbar\omega_{SOP}^i+\hbar^2\omega_{SOP}^i{}^2)$, $t_4=4E_k(E_k-\hbar\omega_{SOP}^i)$,

$t_5 = (E_k + \hbar\omega_{SOP}^i)$, $t_6 = (2E_k^2 + 2E_k\hbar\omega_{SOP}^i + \hbar^2{\omega_{SOP}^i}^2)$, $t_7 = 4E_k(E_k + \hbar\omega_{SOP}^i)$, $t_8 = (2E_k + \hbar\omega_{SOP}^i)$.

**C. Carrier cooling time and heat conductance**

The transport in steady-state measurements is controlled by the electrons near the Fermi level while the transport in high speed devices is determined by electrons whose temperature is raised much above the lattice temperature. Therefore it is also important to understand the temperature dynamics of the hot electrons due to coupling to lattice phonons. Hot carrier dynamics can be probed through experimental techniques such as ARPES [35] and optical differential transmission spectroscopy [6]. The temporal evolution of carrier relaxation, quantified by its electronic temperature $T_e$ is usually described by $\Delta T_e \propto \exp\left(-\frac{t}{\tau^c}\right)$ and can be estimated from the equation [15],

$$\tau^c = C_V \left(\frac{dP}{dT_e}\right)^{-1} \qquad (36)$$

Where $C_V = \frac{d\varepsilon}{dT_e}$ is the electron specific heat, in which $\varepsilon$ is the energy density of graphene. For $T_e \ll \mu/k_B$, $C_V$ shows a linear dependence on $T_e$, i.e $C_V = \frac{8\pi^2}{3}\frac{E_k}{\hbar^2 v_F^2}k_B^2 T_e$ [15]. From the formulae obtained for cooling power we can obtain by direct differentiation the analytical expressions for cooling time for the three phononic modes, which are given as under,

$$\tau_{AP}^C = C_V \frac{\pi^3 \rho v_F^6 \hbar^5}{24 D_{AP}^2 k_B^4 T_e^3}\left(-\left(1+\frac{4}{\pi}\frac{v_p}{v_F}\right)\left\{2k_B T_l[\frac{1}{4}e^{\frac{2\mu}{k_B T_e}} + 8\text{PolyLog}[5,-e^{\frac{\mu}{k_B T_e}}]\right] + \left(\frac{\Delta T}{T}\right)(10k_B T_e - e^{\frac{2\mu}{k_B T_e}}\mu - 16\mu\text{PolyLog}[4,-e^{\frac{\mu}{k_B T_e}}])\left(\frac{1}{4}e^{\frac{2\mu}{k_B T_e}} + 8\text{PolyLog}[5,-e^{\frac{\mu}{k_B T_e}}]\right)\right\}\right)^{-1} \quad (37)$$

$$\tau_{OP}^C = C_V \frac{6\pi\rho\hbar^3 v_F^4}{D_{OP}^2 k_B N_{\omega_{OP}}}\left\{\left(-e^{\frac{\hbar\omega_{OP}}{k_B T_L}}\left[\mu\hbar\omega_{OP}\text{Log}\left(1+e^{\frac{\mu}{k_B T_e}}\right) + \left(\hbar\omega_{OP}^2 + 2k_B T_e(-\mu + \hbar\omega_{OP})\right)\text{PolyLog}\left(2,-e^{\frac{\mu}{k_B T_e}}\right) + 2k_B T_e(3k_B T_e + \hbar\omega_{OP})\text{PolyLog}\left(3,-e^{\frac{\mu}{k_B T_e}}\right)\right] + \right.$$

$$e^{\frac{\hbar\omega_{OP}}{k_BT_e}}\left[\mu\hbar\omega_{OP}\text{Log}\left(1+e^{\frac{\mu}{k_BT_e}}\right)+2k_BT_e\left((-\mu+\hbar\omega_{OP})\text{PolyLog}\left(2,-e^{\frac{\mu}{k_BT_e}}\right)+\right.\right.$$

$$\left.\left.3k_BT_e\text{PolyLog}\left(3,-e^{\frac{\mu}{k_BT_e}}\right)\right)\right]\right\}^{-1} \quad (38)$$

and,

$$\tau_{SOP}^C = C_V \frac{\hbar^4 v_F^3 \epsilon_0 \left(-1+e^{\frac{\hbar\omega_{SOP}^i}{k_BT_l}}\right)}{D_{SOP}^{i\,2} e^{2\hbar\omega_{SOP}^i} e^{-\frac{\hbar\omega_{SOP}^i}{k_BT_e}} k_B} \times \left(\left(e^{\frac{\hbar\omega_{SOP}^i}{k_BT_e}}-e^{\frac{\hbar\omega_{SOP}^i}{k_BT_l}}\right)\mu\text{Log}\left(1+e^{\frac{\mu}{k_BT_e}}\right)+\left(2e^{\frac{\hbar\omega_{SOP}^i}{k_BT_e}}k_BT_e+\right.\right.$$

$$\left.\left.e^{\frac{\hbar\omega_{SOP}^i}{k_BT_l}}(-2k_BT_e-\hbar\omega_{SOP}^i)\right)\text{PolyLog}\left(2,-e^{\frac{\mu}{kb}}\right)\right)^{-1} \quad (39)$$

The heat conductance is defined as [36],

$$G = \frac{dP}{dT_e}\bigg|_{T_e=T_l} \quad (40)$$

The analytical formulae obtained for heat conductance in SLG from the cooling power have been tabulated in Table II.

### III. NUMERICAL AND ANALYTICAL RESULTS AND DISCUSSION

To ascertain the accuracy and validity of the obtained analytical formulae tabulated in Tables I & II we compare it with numerical computation of integral equations, and also with earlier reported results wherever available. For numerical computation of the scattering rates and cooling powers we have used the following values of the parameters, $\rho = 7.6 \times 10^{-8}$ gm/cm$^2$, $v_l = 2.1 \times 10^6$ cm/sec, $v_F = 1 \times 10^8$ cm/sec, $n = 1 \times 10^{12}$ cm$^{-2}$, $D_{AP} = 7.1$ eV, $\hbar\omega_{OP} = 197$ meV, $D_{OP} = 11$ eV/cm, and for graphene on a SiO$_2$ substrate $D_{SOP}^1 = .237 meV, D_{SOP}^2 = 1.612 meV, \hbar\omega_{SOP}^1 = 58.9 meV, \hbar\omega_{SOP}^2 = 156.4 meV$ [15].

### A. Electron-phonon scattering rate

We first consider the case of longitudinal acoustic (LA) phonon scattering and plot the numerical and analytical results from Eqs.(6) and (7) respectively along with the earlier reported analytical result of Eq.(8) in EP regime, as a function of energy and temperature in Figs.2(a) and 2(b), respectively. In Fig.2(a) is plotted the variation of LA phonon scattering rate $(1/\tau)$ with carrier energy $(E_k)$ from four different equations at, $\mu = E_F$ and T=300K. In the four closely lying curves in Fig.2(a) the top curve-A from Eq.(8) represents the approximate quasi elastic scattering, the in-between lying two merged semi-inelastic curves- B & C with and without PB, respectively have been computed from Eq. (7), and the bottom fully numerically evaluated inelastic curve-D from Eq.(5). As can be seen from Table I the relaxation rate in SLG as well as semiconductors is directly proportional to carrier temperature but the energy dependence is linear in graphene whereas for semiconductors it is proportional to square root of carrier energy. The Fig.2(b) shows the log-log plot of the scattering time with temperature at, $\mu = E_F$ and $E_k = 0.5 eV$. Here also we find that the semi inelastic result mimics the same behavior as that of the quasi elastic curve.

| Phonon Mode | Cooling Power $(P_{AP/OP/SOP})$ | Cooling Time $(\tau^C_{AP/OP/SOP})$ | Heat Conductance $(G_{AP/OP/SOP})$ |
|---|---|---|---|
| Acoustic Phonons | $-\dfrac{48\,D_{AP}^2 K_B^5 T_e^5}{\pi^3 \rho v_F^6 \hbar^5}\left(\dfrac{\Delta T}{T_e}\right) \times \left(1+\dfrac{4}{\pi}\dfrac{v_p}{v_F}\right) \times \left(8\text{PolyLog}[5,-e^{\frac{\mu}{k_B T_e}}]+\dfrac{1}{4}e^{\frac{\mu}{k_B T_e}}\mu\right)$ | $C_V \dfrac{\pi^3 \rho\, v_F^6 \hbar^5}{24\, D_{AP}^2 k_B^4 T_e^3}\Bigg(-\left(1+\dfrac{4}{\pi}\dfrac{v_p}{v_F}\right)\Big\{2 k_B T_l[\dfrac{1}{4}e^{\frac{\mu}{k_B T_e}}$ $+\,8\text{PolyLog}[5,-e^{\frac{\mu}{k_B T_e}}]]+(\dfrac{\Delta T}{T})(10 k_B T_e$ $-\,e^{\frac{\mu}{k_B T_e}}\mu$ $-\,16\mu\text{PolyLog}[4,-e^{\frac{\mu}{k_B T_e}}])(\dfrac{1}{4}e^{\frac{\mu}{k_B T_e}}$ $+\,8\text{PolyLog}[5,-e^{\frac{\mu}{k_B T_e}}])\Big\}\Bigg)^{-1}$ | $-\dfrac{24\,D_{AP}^2 k_B^4 T_e^3}{\pi^3 \rho\, v_F^6 \hbar^5}\left(1+\dfrac{4}{\pi}\dfrac{v_p}{v_F}\right)$ $\Big\{2 k_B T_l[\dfrac{1}{4}e^{\frac{\mu}{k_B T_e}}$ $+\,8\text{PolyLog}[5,-e^{\frac{\mu}{k_B T_e}}]]$ $+(\dfrac{\Delta T}{T})(10 k_B T_e - e^{\frac{\mu}{k_B T_e}}\mu$ $-\,16\mu\text{PolyLog}[4,-e^{\frac{\mu}{k_B T_e}}])(\dfrac{1}{4}e^{\frac{\mu}{k_B T_e}}$ $+\,8\text{PolyLog}[5,-e^{\frac{\mu}{k_B T_e}}])\Big\}$ |
| Optical Phonons | $-\dfrac{D_{OP}^2 k_B^2 T_e^2}{6\pi\rho \hbar^3 v_F^4} N_{\omega_{OP}} e^{-\frac{\hbar\omega_{OP}}{k_B T_e}}$ $\times\left(e^{\frac{\hbar\omega_{OP}}{k_B T_l}}-e^{\frac{\hbar\omega_{OP}}{k_B T_e}}\right)\times$ $\Big\{2 k_B T_e \text{PolyLog}[3,-e^{\frac{\mu}{k_B T_e}}]$ $+\,\hbar\omega_{OP}\text{PolyLog}[2,-e^{\frac{\mu}{k_B T_e}}]\Big\}$ | $C_V \dfrac{6\pi\rho\hbar^3 v_F^4}{D_{OP}^2 k_B N_{\omega_{OP}}}\times\Bigg\{\Big(-e^{\frac{\hbar\omega_{OP}}{k_B T_l}}\Big[\mu\hbar\omega_{OP}\text{Log}\Big(1+e^{\frac{\mu}{k_B T_e}}\Big)$ $+\Big(\hbar\omega_{OP}^2$ $+\,2 k_B T_e(-\mu+\hbar\omega_{OP})\Big)\text{PolyLog}\Big(2,-e^{\frac{\mu}{k_B T_e}}\Big)$ $+\,2 k_B T_e(3 k_B T_e+\hbar\omega_{OP})\text{PolyLog}\Big(3,-e^{\frac{\mu}{k_B T_e}}\Big)\Big]$ $+\,e^{\frac{\hbar\omega_{OP}}{k_B T_e}}\Big[\mu\hbar\omega_{OP}\text{Log}\Big(1+e^{\frac{\mu}{k_B T_e}}\Big)$ $+\,2 k_B T_e\Big((-\mu+\hbar\omega_{OP})\text{PolyLog}\Big(2,-e^{\frac{\mu}{k_B T_e}}\Big)$ $+\,3 k_B T_e \text{PolyLog}\Big(3,-e^{\frac{\mu}{k_B T_e}}\Big)\Big)\Big]\Big)\Bigg\}^{-1}$ | $\dfrac{D_{OP}^2 k_B N_{\omega_{OP}}}{6\pi\rho\hbar^3 v_F^4}e^{-\frac{\hbar\omega_{OP}}{k_B T_e}}$ $\Big\{\Big(e^{\frac{\hbar\omega_{OP}}{k_B T_e}}-e^{\frac{\hbar\omega_{OP}}{k_B T_l}}\Big)\Big(\mu\,\hbar\omega_{OP}\text{Log}[1$ $+\,e^{\frac{\mu}{k_B T_e}}]$ $+\,2 k_B T_e(-\mu$ $+\,\hbar\omega_{OP})\text{PolyLog}\Big[2,-e^{\frac{\mu}{k_B T_e}}\Big]\Big)$ $+\Big(-e^{\frac{\hbar\omega_{OP}}{k_B T_l}}\hbar\omega_{OP}^2\text{PolyLog}[2,-e^{\frac{\mu}{k_B T_e}}]$ $+\,e^{\frac{\hbar\omega_{OP}}{k_B T_e}}\text{PolyLog}[3,-e^{\frac{\mu}{k_B T_e}}](2 k_B T_e(3 k_B T_e$ $+\,\hbar\omega_{OP})+3 k_B T_e)\Big\}$ |
| Surface Optical Phonons | $\dfrac{D_{SOP}^{i\;2} e^2 \omega_{SOP}^i k_B^2 T_e^2}{\epsilon_0 v_F^3 \hbar^3} N_{\omega_{SOP}^i} e^{-\frac{\hbar\omega_{SOP}^i}{k_B T_e}}$ $\times\left(e^{\frac{\hbar\omega_{SOP}^i}{k_B T_e}}-e^{\frac{\hbar\omega_{SOP}^i}{k_B T_l}}\right)$ $\times\,\text{PolyLog}[2,-e^{\frac{\mu}{K_B T_e}}]$ | $C_V \dfrac{\hbar^4 v_F^3 \epsilon_0\left(-1+e^{\frac{\hbar\omega_{SOP}^i}{k_B T_l}}\right)}{D_{SOP}^{i\;2} e^2 \hbar\omega_{SOP}^i e^{-\frac{\hbar\omega_{SOP}^i}{k_B T_e}} k_B}$ $\times\Bigg(\bigg(\Big(e^{\frac{\hbar\omega_{SOP}^i}{k_B T_e}}-e^{\frac{\hbar\omega_{SOP}^i}{k_B T_l}}\Big)\mu\text{Log}\Big(1+e^{\frac{\mu}{k_B T_e}}\Big)$ $+\Big(2 e^{\frac{\hbar\omega_{SOP}^i}{k_B T_e}} k_B T_e + e^{\frac{\hbar\omega_{SOP}^i}{k_B T_l}}(-2 k_B T_e$ $-\,\hbar\omega_{SOP}^i)\Big)\text{PolyLog}\Big(2,-e^{\frac{\mu}{kb}}\Big)\bigg)\Bigg)^{-1}$ | $\dfrac{D_{SOP}^{i\;2} e^{-\frac{\hbar\omega_{SOP}^i}{k_B T_e}} e^2 k_B \hbar\omega_{SOP}^i}{\hbar^4 v_F^3 \epsilon_0\left(-1+e^{\frac{\hbar\omega_{SOP}^i}{k_B T_l}}\right)}$ $\Big\{\Big(e^{\frac{\hbar\omega_{SOP}^i}{k_B T_e}}$ $-\,e^{\frac{\hbar\omega_{SOP}^i}{k_B T_l}}\Big)\Big(\mu\,\text{Log}\Big(1+e^{\frac{\mu}{k_B T_e}}\Big)$ $+\,2 k_B T_e\Big)$ $-\,e^{\frac{\hbar\omega_{SOP}^i}{k_B T_l}}\hbar\omega_{SOP}^i\Big)\text{PolyLog}\Big(2,-e^{\frac{\mu}{k_B T_e}}\Big)\Big\}$ |

TABLE II. Summary of analytical results for cooling power $(P_{AP/OP/SOP})$, cooling time $(\tau^C_{AP/OP/SOP})$ and Heat conductance $(G_{AP/OP/SOP})$ through acoustic, optical and surface polar optical phonon scattering in SLG.

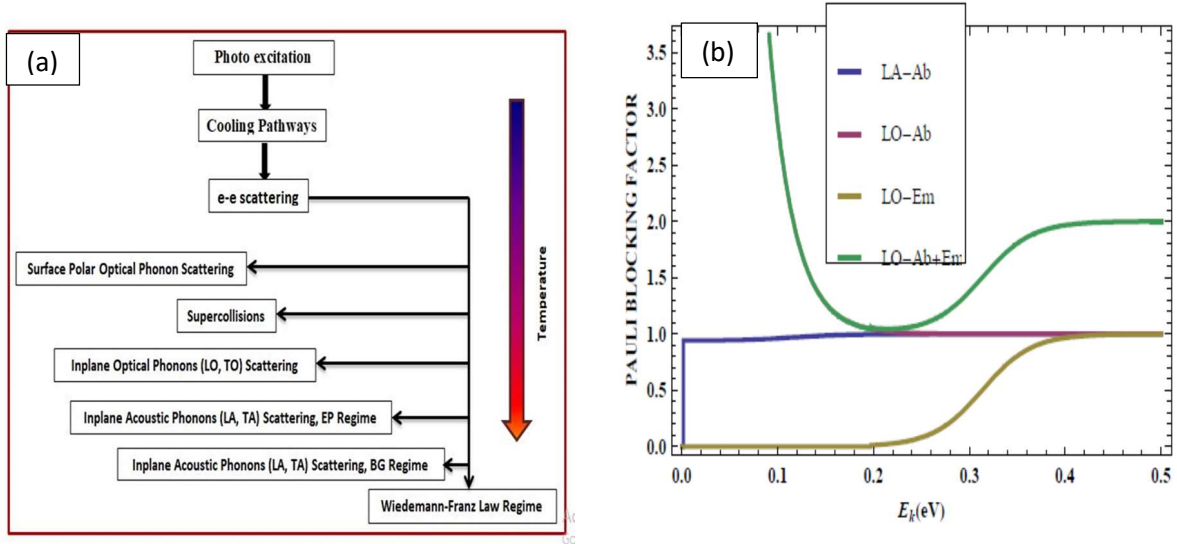

FIG.1. (a) Illustration of major cooling pathways of photoexcited electrons in graphene on a substrate, (b) Pauli blocking factor vs Electron energy at T=300K in which curves for LA Absorption process, LO- Absorption and Emission Processes separately, and LO-sum of absorption and emission processes .

From both the Figs.2(a) & 2(b) it can be concluded that the approximate semi-inelastic EP limit LA phonon scattering analytical and numerical results as a function of energy and temperature do not differ from the earlier reported quasi elastic LA phonon analytical result of Eq.(8), which means that the elastic scattering approximations made in obtaining the EP limit results for estimating the scattering rate holds good as the LA phonon energy and PB factor's contribution to the scattering rate is negligibly small. We have not shown the behavior of scattering rate through transverse acoustic TA mode as the functional dependence with temperature and energy remains the same with only a replacement of the value of velocity in the equations.

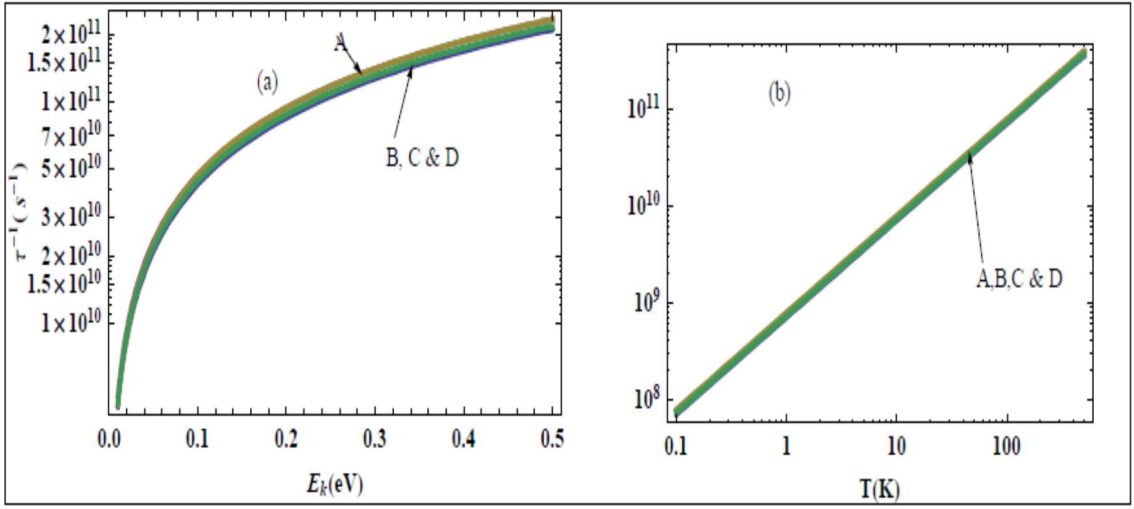

FIG.2. Scattering rate, $(1/\tau)$ as a function of electron energy, $(E_k)$ and Temperature $(T_e)$ for LA phonon scattering at $\mu = E_F$. The Fig. 2(a) is at $T_e$=300K. in which the top curve-A from Eq.(7) represents approximate semi inelastic scattering with PB, the in-between lying two merged semi inelastic curve-B from Eq.(7) without PB and fully numerical inelastic curve-C from Eq. (6) with PB, and the bottom curve-D from the quasi elastic analytical expression Eq.(8). The Fig. 2(b) is log-log plot at $E_k = 0.5 eV$.

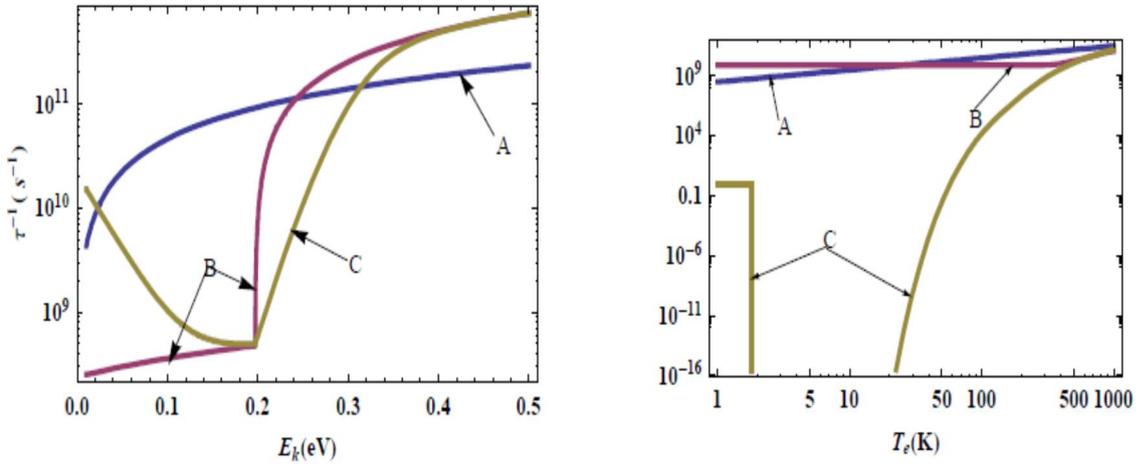

FIG.3. Scattering rate, $(1/\tau)$ as a function of electron energy, $(E_k)$ and Temperature $(T_e)$ for LA phonon scattering (curve-A) and optical phonon scattering (curves-B & C) at , $\mu = E_F$. The Fig.3(a) is a semi-log plot at $T_e$=300K in which the LA curve-A is computed from Eq. (7) and the LO curves-B & C are computed from Eqs.(11) and (10), respectively. The Fig.3(b) is log-log plot at $E_k = 0.2 eV$. In the curves-B & C the relaxation rate before $\hbar\omega_{OP} = 0.197\ meV$ denotes scattering by LO phonon absorption and afterwards it is due to both absorption and emission of LO phonons.

The Figs.3(a) and 3(b) demonstrate the effects of the scattering by longitudinal optical (LO) phonons and its comparison with the LA phonon scattering. In the two curves the scattering rate, $(1/\tau)$ as a function of electron energy, $(E_k)$ and temperature $(T_e)$ for LO and LA phonon scattering have been plotted at, $\mu = E_F$. In both the plots, the curve-A depicts the LA scattering rate and curves -B & C represent the LO phonon scattering rates. The Fig. 3(a) is a semi log plot at $T_e$=300K in which the LA curve-A is computed from Eq. (7) and the LO curves-B & C are computed from Eq.(11) without PB and from Eq.(10) with PB, respectively. The Fig. 3(b) is a log-log plot at $E_k = 0.2eV$. In the case of LO scattering rate shown by curves-B & C the effect of PB around the LO phonon energy, $\hbar\omega_{OP} = 0.197\ meV$ is clearly noticeable. The relaxation rate before $\hbar\omega_{OP} = 0.197\ meV$ represents scattering by LO phonon absorption and afterwards it is due to both absorption and emission of LO phonons. It can be noticed from Eqs. (8) &(11) for LA and LO phonon scattering rates without PB, that the scattering rate varies linearly with energy. This is because the DOS in graphene is a linear function of energy and as a result, the number of available final states for electrons increases with energy away from the Dirac point. But the LO phonon relaxation rate considering PB factor (curve-C) shows considerable difference as compared to the approximate curve-B, as the linear energy function is modified by exponential energy dependent factors. This clearly proves that in the optical scattering case the PB factor cannot be ignored unlike the case of acoustic scattering where it was insignificant. The different behavior of the PB curve can be explained from the PB factor, $\frac{1-f(E_{k'})}{1-f(E_k)}$ whose numerical value for emission plus absorption attains the value ($\leq 2$) at $E_k = E_F$ for $T_e \geq 0K$, but the value decreases with increase in energy as can also be seen from Fig.1(b) for only absorption LO process, as the states begin to be increasingly occupied resulting in the continuous decrease of the PB factor till it reaches its minimum value which is zero at $T_e = 0K$ and $E_k = \hbar\omega_{OP}$. (Note here the graph is plotted at $T_e = 300K$ hence the scattering rate shows a finite value at threshold energy). Therefore the curve-C declines in the absorption region $E_k \leq \hbar\omega_{OP} = 0.197\ meV$. After the LO phonon energy threshold $E_k \geq \hbar\omega_{OP}$, emission and absorption both takes place with an increase in the PB factor and the two curves with and without PB merge in the non-degenerate regime as the Pauli blocking is lifted. The asymmetry observed in the curve-C around the LO phonon energy is because on the LHS side of the optical threshold there is contribution from only one process while on the RHS of threshold there are two processes of absorption and emission taking place, which results in a slightly steeper incline in contrast to the slow decline on the absorption side. However as compared to curve-B with no PB factor there is seen suppression of scattering rate in curve-C because of PB, even when electrons have higher energy than the threshold phonon emission energy. Also the LO phonon scattering curves- B & C are observed to dominate over the LA scattering curve-A soon after the optical energy threshold. Though the LO phonon energy scales over the room temperature energy still LO phonons serve as an important source of total momentum relaxation below room temperature [37-38]. The Fig.3(b) plotted at $E_k = 0.2 eV$ corroborates the observations of Fig.3(a). The LO phonon scattering with PB in the absorption region shows a step function behavior as the PB factor is almost equal to one at $E_k = E_F$ at low temperatures, but at higher temperatures the PB factor begins to contribute and together with the phonon emission process increase the scattering rate. The LA and LO phonon scattering without PB in Fig.3(b) stays above the LO scattering without PB curve at $E_k = 0.2 eV$ at low values of temperature and approaches the other two curves only at high temperature values, as also seen from Fig.3(a). The comparison of our result with that investigated using density-functional perturbation theory is deferred to the end part of this subsection where we compare the scattering rates of all the three phononic modes.

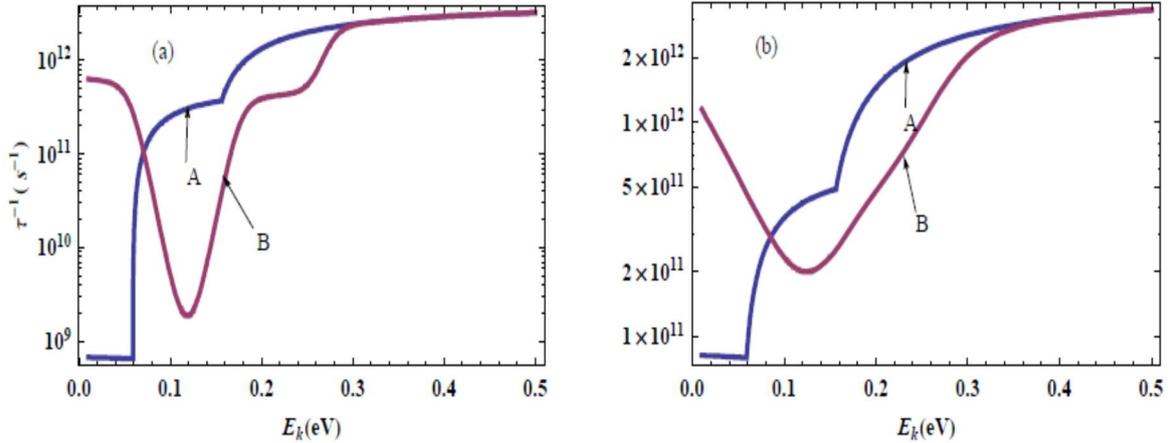

FIG.4.  Scattering rate, $(1/\tau)$ in SLG on SiO$_2$ as a function of electron energy, $(E_k)$ for SOP phonon at, $\mu = E_F$ at $T_e = 100K$ & $T_e = 300K$ in Figs.4a & 4b, respectively. The merged curves-A & B are with PB factor in which curve-A (without $\exp(-qd)$) is analytical curve from Eq.(13) and numerical curve-B is with $\exp(-qd)$ from Eq.(12). The merged curves-C & D are without PB factor in which curve-C (without $\exp(-qd)$] ) is analytical curve from Eq.(14) and numerical curve-D is with $\exp(-qd)$ from Eq.(12). In the curves-C & D, the relaxation rate before $\hbar\omega_{SOP} = 0.058\ eV$ denotes scattering by SOP absorption and afterwards it is due to both absorption and emission. The Fig. 4(b) is log-log plot at $T_e = 300K$.

The inelastic SOP scattering has been touted to play a prominent role both at low and high field mobility for graphene on a substrate, and in case of low field bias the extrinsic SOP or intrinsic LO scattering determines the current saturation, whereas the LA scattering determines the low-field mobility [39]. The SOP phonon scattering rate, $(\tau_{SOP}^{-1})$ in SLG as a function of electron energy, $(E_k)$ at $\mu = E_F$ is shown in Figs. 4(a) and 4(b). The merged curves-A & B in Fig.4(a) at T=300K are with PB factor in which curve-A without $\exp(-qd)$, is analytical curve from Eq.(13) and the numerical curve-B with $\exp(-qd)$ is from Eq.(12). The merged curves-C & D are without PB factor in which curve-C without $\exp(-qd)$ is analytical curve from Eq.(14) and numerical curve-D is with $\exp(-qd)$] from Eq.(12). In the curves-C & D, the relaxation rate before $\hbar\omega_{SOP} = 0.058\ eV$ denotes scattering by SOP absorption and afterwards it is due to both absorption and emission of SOP. In SOP scattering the effect of PB is also significant like the case of optical phonons in Fig.3. Also in case of SOP the kink observed in PB curves-A & B happen at higher energy than the threshold as compared to LO case, which is due to the fact that though the SOP emission process begins at the threshold energy $\hbar\omega_{SOP}$ but the PB factor continues to decline because $E_F > \hbar\omega_{SOP}$ whereas in LO scattering $E_F \sim \hbar\omega_{OP}$ and the decline reaches its minimum at the threshold value itself. The SOP plotted curves from the obtained analytical equations are in very good agreement with that obtained through a Monte Carlo simulation, and for $E_k \geq 2.9 eV$ both the calculations produce a scattering rate of $\tau_{SOP}^{-1}=10^{13}$ sec. [40]. The Fig. 4(b) is log-log plot at $E_k = 0.2 eV$ that also confirms the enhanced PB effect in SOP scattering.

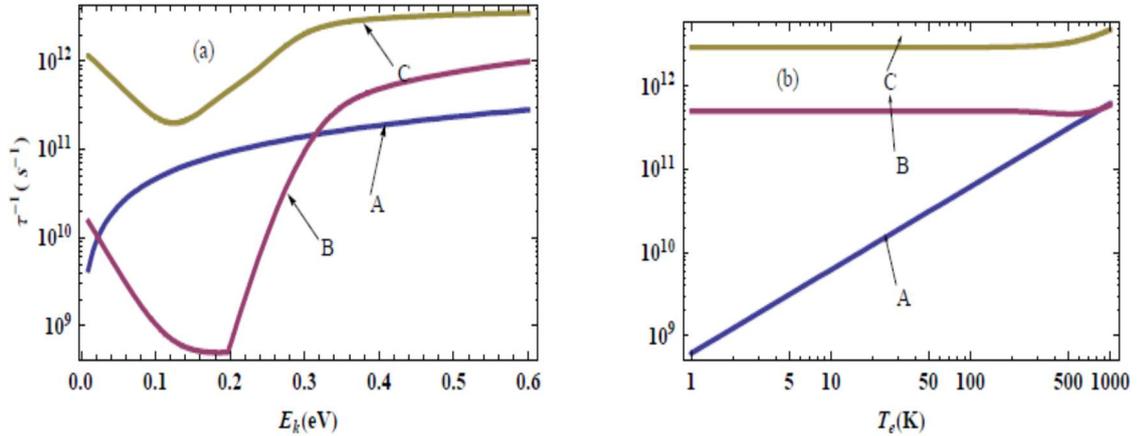

FIG.5. Comparative scattering rate, $(1/\tau)$ in SLG as a function of electron energy, $(E_k)$ and Temperature $T_e$ for acoustic, optical and SOP at $\mu = E_F$. The curve A is for acoustic scattering, B is for optical phonon and C is for SOP. The Fig. 5(a) is a semi-log plot at $T_e$=300K and Fig. 5(b) is a log-log plot at $E_k = 0.4 eV$.

The Fig.5 illustrates the comparative behavior of the three scattering rates due to LA, LO and SOP phononic modes. Here we observe that at higher energies the LO phonon scattering dominates while the LA phonon scattering too becomes significant as compared to SOP phonon scattering which shows a nearly constant energy dependence except near the surface phonon energy. The power loss is found predominantly to be due to LA phonon emission below ~200–300 K with LO phonon emission taking over as the dominant energy loss mechanism above this temperature range. This is also indicated in a study where high-energy optical and zone-boundary phonons in graphene $\hbar\omega_{SOP} = 0.150\ eV$ are responsible for 50% of resistivity due to electron-phonon scattering even at room temperature and becomes dominant at higher temperatures [41]. The LO and LA scattering rates obtained from our analytical results are also consistent with that reported using density-functional theory, as for $E_k > 0.3\ eV$ the value of scattering rate touches, $\tau_{OP}^{-1}=10^{12}$ sec$^{-1}$ , and LA scattering rate differs rate by an order of magnitude less than LO rate, $\tau_{AP}^{-1}=10^{11}$ sec [42]. In general we conclude from this discussion that in SLG the SOP scattering dominates the scattering due to the other two modes at all values of energies and temperatures, and that the LO phonons provides the dominant scattering process at high electron energies in comparison to LA phonons. The LO phonons are much more effective than LA phonons in scattering electrons at high electron energies (~>0.3eV), whereas SOP phonons contribution to scattering rate is more than the LA phonons at low energies (~<0.12eV). In summary the analytical results obtained correctly reproduce the numerical results and also substantiate the earlier finding that all in-plane phonons play an important role in electron phonon interactions in SLG and must be taken into account for transport studies at room temperature [42].

**B. Cooling power**

In the following we discuss the analytical and numerical results for the hot electron temperature $T_e$ dependence of the cooling power $P(T_e)$ due to acoustic, optical and SOP scattering. The estimation of cooling power rate is not only important for understanding the hot electron dynamics but it also provides another method of determining the e-p coupling constant. We asssume a common lattice temperature for all the three phononic modes. For numerical computation of the cooling rate we have used the same values of parameters as done in the estimation of scattering rate however few other additional parameters required in cooling rate computation are $T_l = 10\ K$, $v_s = 1.7 \times 10^6$ cm/sec, where $v_s$ is effective sound velocity in graphene given by $2v_s^{-2} = v_l^{-2}+v_t^{-2}$ , in which $v_l=2.4 \times 10^6$ cm/sec and $v_t = 1.4 \times 10^6$ cm/sec. [15].

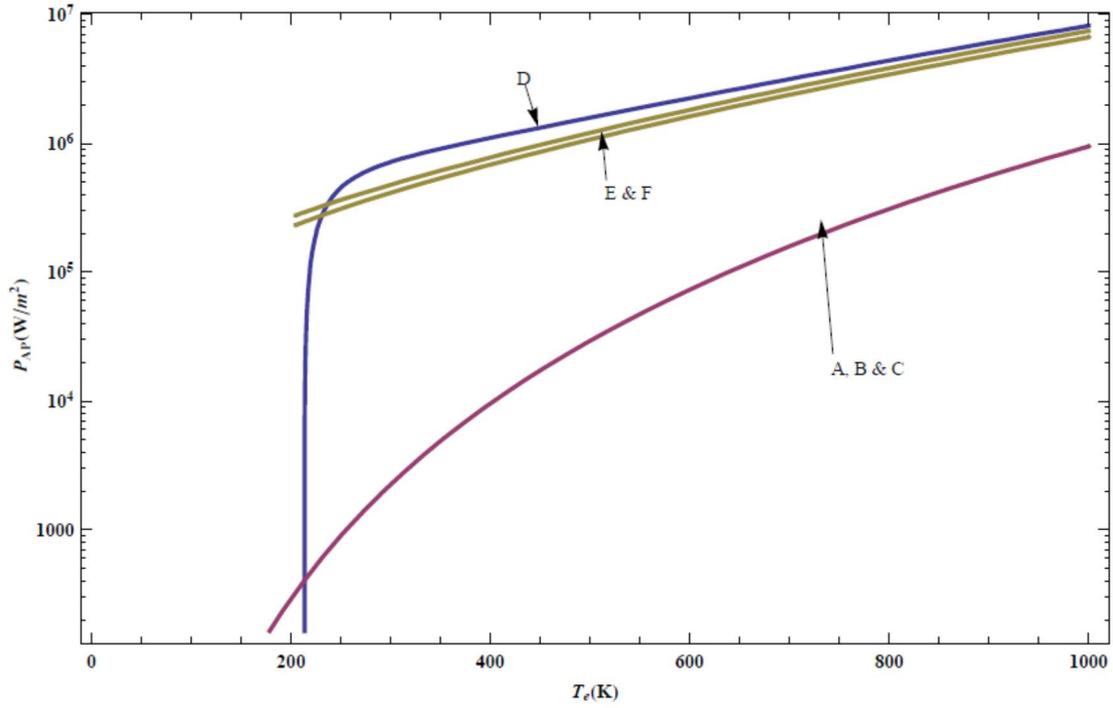

FIG.6. Cooling power density $P_{AP}(W/m^2)$ in SLG as a function of temperature, $T_e$ for acoustic phonon scattering at $T_l = 10K$. Overlapping curves-A, B & C are approximate analytical & numerical (Ref.15) while curve-C is full numerical from Eq.(19) at , $\mu = 0$. The curves-D, E & F are the same curves-A, B & C plotted at $\mu = 0.2 eV$.

The cooling power per unit area as a function of the hot electron temperature $T_e$ for unscreened acoustic DP coupling at $\mu = 0$ is shown in Fig. 6. The overlapping curves-A, B & C are approximate analytical & numerical from Eqs.(22) and (23), respectively, while curve-C is full numerical result from Eq.(19) at , $\mu = 0$. The curves-D, E & F represent the same curves-A, B & C but at $\mu = 0.2 eV$. The merged curves-A, B & C clearly show that the derived analytical result from Eq.(22) at $\mu = 0$ perfectly reproduces the cooling rate from the integral representation given by Eq. (23) [15] and very fairly matches the numerical results at $\mu = 0.2 eV$. This also proves that the approximation done in obtaining integral equation for acoustic mode in Ref.[15] are very fairly accurate.

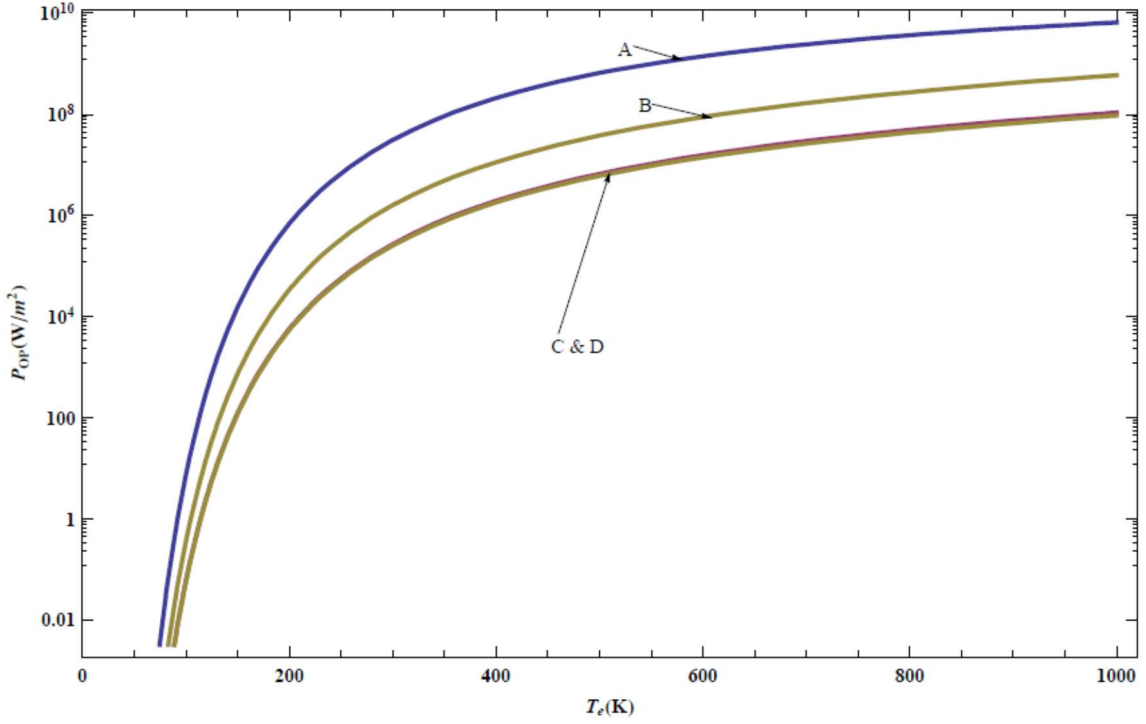

FIG.7. Cooling power density $P_{OP}(W/m^2)$ in SLG as a function of electron temperature, $T_e$ for optical phonon scattering at, $\mu = 200 meV$. The top curve- A is from approximate analytical result from Eq.(9) reported in Ref.[13], while the in-between lying curve-B is the complete numerical expression from Eq.(24). the bottom overlapping curves- C & D are obtained approximate analytical result and approximate numerical results respectively from Eqs.(28) and (26), respectively.

Next we consider the optical phonon scattering and show the computed results of the optical cooling rate $P_{OP}$ in Fig.7., at $\mu = 200 meV$. Since in the case of optical phonons an empirical formulae is available in the literature therefore we compare our obtained approximate analytical result with the reported formula in Ref.[13], besides comparing with the approximate numerical and complete numerical results in this paper. The top curve-A is from approximate analytical result from Eq.(29) reported in Ref.[12-13], while the in-between lying curve-B is the complete numerical expression from Eq.(24), the bottom overlapping curves- C & D are the obtained approximate analytical and approximate numerical results from Eqs.(28) and (26), respectively. This again shows that the obtained closed form Eq.(28) exactly determines the solution of integral cooling rate Eq. (26). Though our obtained analytical solution marginally underestimates the complete numerical result but still it is closer to it than the reported analytical result in Ref.[12], which overestimates it by a greater amount. Though the larger value of optical phonon energy ($\hbar\omega_{OP} = 0.197\ eV$) suppresses the optical phonon contribution to electron-lattice relaxation below a few hundred kelvin but we observe a significant contribution of optical phonon scattering to the $P_{OP}(T_e)$ even at room temperature, which was earlier termed insignificant [8].

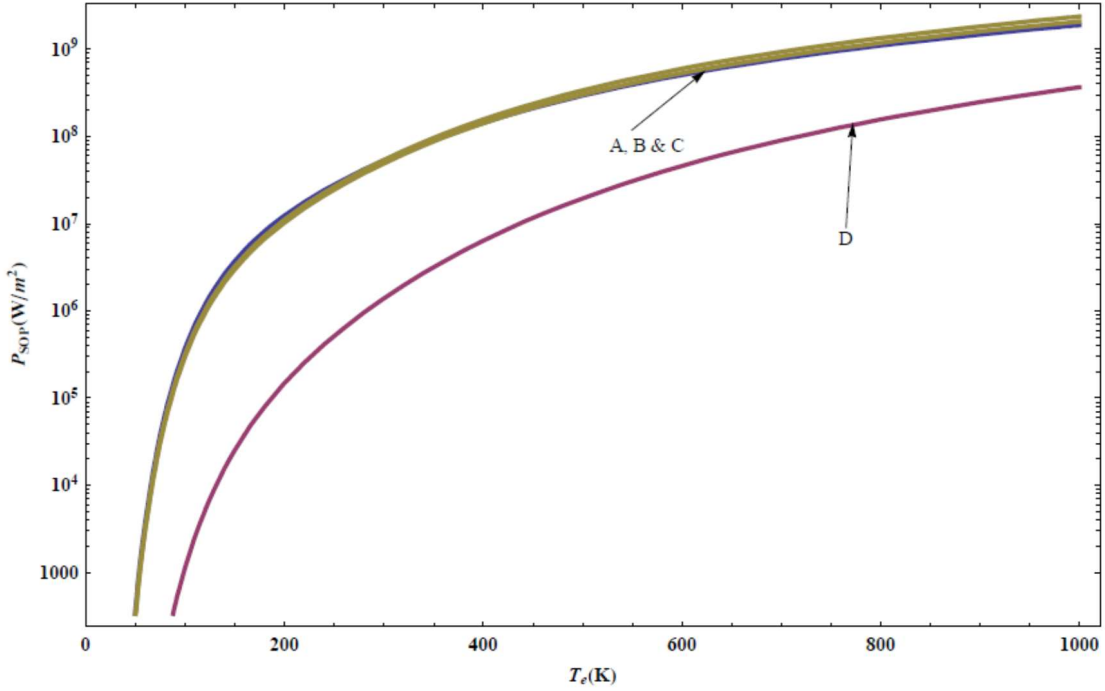

FIG.8. Cooling Power density $P_{SOP}(W/m^2)$ as a function of temperature, $T_e$ for intraband unscreened SOP scattering. The merged curves-A & B respectively are the approximate analytical and complete numerical results with, and without substrate distance factor, at $\mu = 200meV$. The curve-D is analytical curve at, $\mu = 0meV$.

The numerical and analytical results for intraband cooling rate $P_{SOP}$ of a SLG on a substrate due to unscreened SOP is shown in Fig.8 for two values of $\mu = 200meV$ and $\mu = 0meV$. When the results are computed at $\mu = 200meV$ we observe that the analytical curve-A from Eq.(35), the complete numerical curves-B & C from Eq.(31), first with and then without the substrate distance factor, all merge together. The curve-D is analytical curve plotted at $\mu = 0meV$. From the computed results in Fig. 8 we can say that overall the analytical result fairly produces the numerical result. We further compare the obtained SOP analytical result of Eq.(35) with the optical and acoustical analytical cooling rates from Eqs. (22), (28), respectively represented by curves A, B and C in Fig.9 from the three phonon coupling mechanism. We also include the disorder assisted cooling rate reported in Ref.[10],

$$P_{AP;Dis} = \frac{14.43\, D_{AP}^2 \mu^2 k_B^3 T_e^3}{\rho \hbar^5 v_F^4 v_p^2 k_F l} \qquad (41)$$

shown in the plot at $k_F l = 20$, by curve-D. The SOP cooling rate dominates all the cooling rates due to the three other phononic modes at all temperatures. The SOP is followed by cooling rate due to super-collisions. The optical cooling rate overtakes the acoustic cooling rate at~ 300K for $\mu = 100 meV$ and competes with supercollision cooling at higher temperatures but lag behinds $P_{AP}$ at lower temperatures, and the disorder assisted cooling dominates the other two optical and acoustic modes at low temperatures, with the optical phonon mode beginning to contribute equally to the cooling rate at elevated temperatures. The acoustic phonon cooling rate though remains suppressed by over two orders of magnitude as compared to $P_{SOP}$ and by over an order of magnitude as compared to $P_{DIS}$.

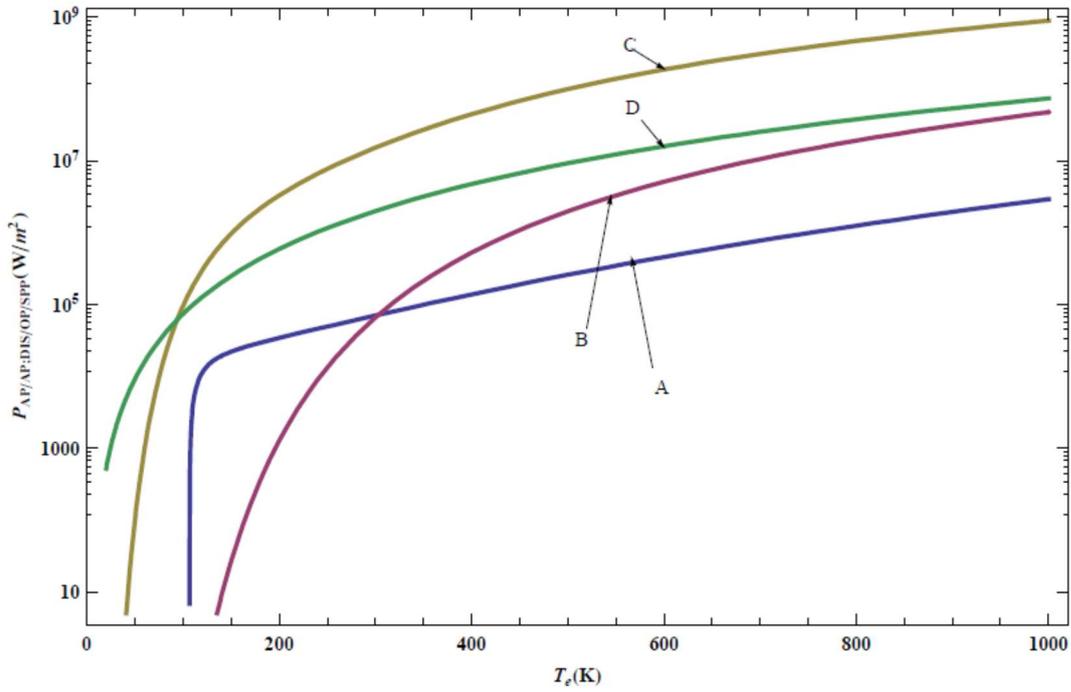

FIG.9. Comparative analytical estimates for the cooling power density $P_{AP/OP/SOP/AP;DIS}(W/m^2)$ in SLG as a function of electronic temperature $T_e$ due to acoustic, optical, SOP and acoustic phonon supercollision scattering processes at $\mu = 100 meV$.

### C. Carrier cooling time and heat conductance

Finally we estimate the carrier cooling time and heat conductance from our obtained unscreened analytical formulae tabulated in Table II and exhibit their plots at $\mu = 0 meV$ in Figs.10 (a) and 10(b), respectively. In both the Figs.10 (a) and 10(b), curve-A represents the effect due to acoustic phonons while the curve-B depicts optic phonon scattering and the curve-C shows SOP scattering. It is observed from Fig.10(a), that at higher temperatures the cooling time is fastest ($10^{-15}$sec) due to SOP scattering followed by optical scattering with a drop of an order of magnitude in time ($10^{-14}$sec), which more or less prevails starting from 1000K to 200K. At high temperatures the slowest relaxation ($10^{-13}$sec. at 1000K), is due to acoustic phonon scatterings which is about two orders of magnitude smaller than SOP scattering, and an order of magnitude lesser than optic phonon scattering. However the margin between the two starts decreasing from ~400K overtaking the optic curve at 200K and equalling the SOP cooling time at about 10K. This means in case of graphene on a substrate the dominating SOP scattering plays an important role at all temperatures and that the acoustic scattering too becomes substantial down from 200K, and that the optic scattering from above 200K prevails over acoustic scattering. The cooling time obtained from the unscreened cooling power density formulae somewhat overestimates the cooling time as reported from numerical computation [15]. This we feel is due to the use of the analytical formula for specific heat in the estimation of cooling time at $\mu = 0 meV$, and $C_V = \frac{8\pi^2}{3} \frac{E_k}{\hbar^2 v_F^2} k_B^2 T_e$ when $T_e \ll \mu/k_B$ is not valid in the in the EP regime.

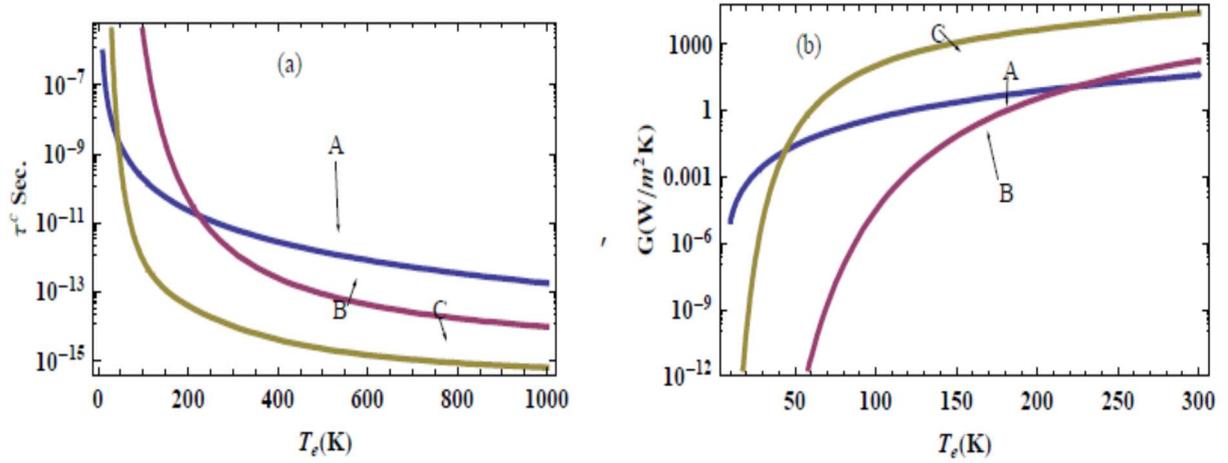

**FIG.10.** Carrier's cooling time, $\tau^c$ and Heat conductance, $G$ plotted as a function of temperature, $T_e$ for acoustic, optic and SOP scattering at $\mu = 0 meV$.

The Fig.10(b) shows the variation of heat conductance from the listed formulae in Table II for the three phononic modes calculated from Eq.(40). Here also the trend is similar to that observed in Fig.9 for the case of cooling power density with the leading contribution by SOP, followed by optical and then the acoustic phonons. The order of the magnitude for acoustic and optics modes in general matches with that reported in Ref.[36] for supercollision enabled multi phonon acoustic and optic scattering processes dependent cooling, where between 200-300 K, $G_{AP}$ lies between $= (1 - 10^2)\frac{W}{m^2 K}, G_{OP}$. The heat conductance due to SOP is found to be highest that is $G_{SOP}$ lying between $(10^3 - 10^4)\frac{W}{m^2 K}$ for 100-300 K. So the analytical formulae for cooling time and heat conductance fairly produce the reported results. In summary all the analytical results obtained correctly reproduce the numerical results and also substantiate the earlier finding that all in-plane phonons play an important role in electron phonon interactions in SLG and must be considered for transport studies at room temperature [41]. It is very much expected that this study will boost attempts to obtain analytical solutions for transport quantities in other Dirac systems and in bilayer graphene.

## IV. CONCLUSIONS

We revisited the problem of inelastic scattering and cooling of photoexcited electrons through coupling with acoustic, optical, and surface polar optical phonons and obtained analytical results for all the cases in place of their earlier reported integral representations. Our analytical results for scattering are in good agreement with the corresponding earlier reported integral representations. We also considered the effect of Pauli blocking on the inelastic scattering and cooling rates and found that the effect of Pauli Blocking is more pronounced for optical and surface polar phonon scattering while it is negligibly small for acoustic phonon scattering rate. Our study shows that the LO phonons in SLG provides the dominant scattering mechanism process at high electron energies (~>0.3eV) and above this limit, it is much more effective than LA phonons scattering, whereas SOP phonons contribution to scattering rate is more than the LA phonons at low energies (~<0.12eV).

In the case of cooling power density, we obtained analytical results due to all the three phononic modes considered in this study and find that the SOP cooling rate dominates all the cooling rates due to the three other phononic modes at all temperatures, including disorder assisted acoustic phonon scattering. The SOP is followed by a cooling rate due to super-collisions. The optical cooling rate overtakes the acoustic cooling rate at~ 300K for $\mu$=100 meV and competes with supercollision cooling at higher temperatures but lag behinds $P_{AP}$ at lower temperatures, and the disorder assisted cooling dominates the other two optical and acoustic modes at low temperatures, with the optical phonon mode beginning to contribute equally to the cooling rate at elevated temperatures. The acoustic phonon cooling rate though remains suppressed by over two orders of magnitude as compared to $P_{SOP}$ and by over an order of magnitude as compared to $P_{DIS}$. Further from our obtained analytical formulae for the cooling rates, we could deduce analytical formulae for cooling time and heat conductance which also produce results in good agreement with earlier reported numerical and simulation-based methods. We believe that this study will enable further investigations to obtain analytical solutions for transport quantities in other Dirac systems and in bilayer graphene.